\documentclass[prb,twocolumn,showpacs]{revtex4}
\usepackage{color,soul}
\usepackage{amsmath}
\usepackage{dcolumn}
\usepackage{graphicx}
\usepackage{bm}
\usepackage{amssymb}
\usepackage{subfigure}
\usepackage{diagbox}
\begin{document}

\title{Disorder effect on the superfluid density and the origin of the pseudogap end point in the cuprate superconductors}
\author{Rong Cheng}
\affiliation{School of Physics, Renmin University of China, Beijing 100872, P.R.China}

\author{Tao Li}
\email{litao_phys@ruc.edu.cn}
\affiliation{School of Physics, Renmin University of China, Beijing 100872, P.R.China}

\author{Jianhua Yang}
\affiliation{School of Physics, Ningxia University, Yinchuan 750021, P.R.China}

\begin{abstract}
A major puzzle in the study of the cuprate superconductivity is the origin of the pseudogap end point, around which dramatic transition in electronic properties of the system have been found. Intriguingly, such a critical doping is also where the superfluid density of the system reaches its maximum. A non-monotonic doping dependence of the superfluid density is rather unusual since the Drude weight of the cuprate system is found to increase monotonically with the doping concentration. It is generally believed that such a peculiar behavior should be attributed to both the strongly correlated nature of the cuprate system and the disorder effect. However, the calculation of the superfluid density in a disordered and strongly correlated model is a formidable task. In this work, we develop a variational theory for the zero temperature superfluid density of the disordered $t-J$ model. This is achieved in two steps. First, we perform an unrestricted variational optimization of an RVB variational ground state for the disordered $t-J$ model. Second, we construct the variational state that describes the paramagnetic current response on such an RVB state. The zero temperature superfluid density $\rho_{s}(0)$ is then extracted from the curvature of the variational ground state energy of the system as a function of the external electromagnetic field. We find that $\rho_{s}(0)$ computed in this way is remarkably robust against the disorder effect. More specifically, we find that $\rho_{s}(0)$ is a monotonically increasing function of doping concentration $x$ and scales linearly with the total optical weight. This is consistent with the observation in the underdoped cuprates but is strongly at odd with the behavior in the overdoped cuprates. The strong contrast between the disorder effect in the underdoped and the overdoped regime lends strong support to our previous proposal that there exist a Mott transition between a doped-Mott-insulating metal in the underdoped regime and a fermi-liquid-like metal in the overdoped regime around the pseudogap end point.         
\end{abstract}

\maketitle

\section{Introduction}
A major puzzle in the study of the superconductivity in the high-T$_{c}$ cuprates is the origin of the pseudogap end point. Dramatic transition of electronic properties have been found around such a critical doping\cite{Proust,Hussey} which will be denoted as $x^{*}$ in the following. For example, ARPES measurements find that the peak-dip-hump structure in the anti-nodal spectrum is suddenly suppressed around $x^{*}$\cite{Chen}. It is also found that the critical fan of the spectacular strange metal behavior in the normal state is rooted from such a special doping, where a sharp logarithmic peak in the specific heat is reported in thermodynamic measurement\cite{Taillefer,CV}. At the same time, a dramatic change from $n_{H}\approx x$ to $n_{H}\approx 1+x$ doping dependence of the measured Hall number is also reported across $x^{*}$\cite{Hall3}. While all these phenomenologies imply a quantum critical point of some kind at $x^{*}$, such a doping is far away from any known symmetry breaking phase transition in the cuprate superconductors\cite{Hussey}.   

Intriguingly, the pseudogap end point is also found to be the doping where the superfluid density of the system reaches its maximum\cite{Hussey,Tallon,Uemura,Bozovic}. This becomes especially clear in the zinc-doped cuprates, in which the superconductivity is deliberately suppressed by introducing such a strong in-plane scatters. Measurements find that superconductivity survives the most robustly against zinc doping at $x^{*}$. A non-monotonic doping dependence of the superfluid density is rather unusual since we know that effective carrier density increases monotonically with doping in such a doped Mott insulator. This is supported by the monotonic doping dependence of the measured Drude weight in the optical conductivity spectrum\cite{Michon}. In recent years, it is claimed that the unexpected decrease of the superfluid density with doping for $x>x^{*}$ should be attributed to the disorder effect in a d-wave BCS superconductor\cite{Hone1,Hone,Lizx,Kivelson,Broun,Berg}. More specifically, while non-magnetic impurity is expected to have little effect on superconductivity in a conventional s-wave BCS superconductor, it can act as a strong pair breaker in a d-wave BCS superconductor as a result of the non-trivial phase structure of pairing potential of such a system\cite{Markowitz,Alloul,Atkinson,Xiang,Atkinson1,Anderson}. Indeed, calculations based on a dirty-d-wave BCS scenario find that a significant part of the Drude weight remains uncondensed even at zero temperature. This behavior is indeed observed in the heavily overdoped cuprates\cite{Mahmood}. The predicted strong spatial modulation of the pairing gap in the presence of disorder is also qualitatively in agreement with recent STM measurement in strongly overdoped cuprate systems\cite{Tromp}.

Given the fragility of the d-wave pairing to disorder effect in the overdoped cuprates, it seems puzzling why the same d-wave superconducting order in underdoped regime is so robust against disorder effect, which is unavoidable in a doped Mott insulator as the cuprate superconductor. According to the Uemura's plot\cite{Uemura}, the superfluid density scales linearly with the doping concentration in the underdoped regime. This is just the intrinsic behavior that we expect for a doped Mott insulator. In fact, as a result of poor screening in the underdoped regime, the strength of disorder effect caused by the out-of-plane dopant is in some sense even stronger in the underdoped regime. Some researchers attribute such a counter-intuitive behavior to the strong correlation effect, more specifically, the non-BCS character of the electron pairing in the underdoped regime. For example, from a Bogoliubov-de Gennes-type mean field calculation supplemented with Gutzwiller approximation(B-dG+GA) on the no double occupancy constraint in the $t-J$ model, the authors of Ref[\onlinecite{Grag}] argued that the effect of the disorder potential on a given site is compensated by the adjustment of the Gutzwiller factor for the kinetic energy term that connects to this site. The effect of the disorder potential can thus be healed on a much shorter length scale. The obtained quasiparticle density of state(DOS) in the presence of the disorder potential is found to be remarkably homogeneous. However, it is unclear if the crude Gutzwiller approximation can capture such a subtle correlation effect\cite{Grag,Chakra,Ghosal}. In particular, the predicted DOS from such a BdG+GA treatment seems to be too homogeneous to be consistent with observation in underdoped cuprates.

Recently, we have performed a systematic variational study on the disorder effect in the $t-J$ model within the resonating valence bond(RVB) theory framework\cite{Yang}. We find that the d-wave pairing in the system is remarkably more robust against the disorder potential than that in a conventional d-wave BCS superconductor. In particular, the d-wave phase structure of the pairing potential is preserved both locally and globally in rather strong disorder potential. This is very different from the situation in a disordered d-wave BCS superconductor, in which the frustration in the phase of the pairing potential can be so strong that even the time reversal symmetry can be spontaneously broken\cite{Lizx}. However, we do notice a key difference with the prediction of the BdG+GA calculation. More specifically, while the phase structure of paring potential remains intact, the magnitude of the pairing potential obtained from our variational optimization exhibits substantial spatial modulation in the presence of the disorder potential. We find that such spatial modulation can be understood as the secondary effect of the modulation in the local hole concentration. 

We also find that the remarkable robustness of the d-wave pairing in the disordered $t-J$ model can be attributed to the spin-charge separation mechanism in the RVB picture, through which the d-wave RVB pairing of the charge neutral spinons becomes essentially immune to the disorder potential, except for the secondary effect related to the modulation of the local doping level by the disorder potential. Based on these results, we propose that there exists a Mott transition at the pseudogap end point $x^{*}$, where the RVB pairing in the underdoped regime is transmuted into the increasingly more BCS-like pairing for $x > x^{*}$, whose increasing fragility against the disorder effect leads to the ultimate suppression of superconductivity around the end point of the superconducting dome. While such a scenario is rather appealing, the quantities calculated in the above variational study, namely, the pairing potential and the off-diagonal-long-range-order(ODLRO), are not accessible to measurement. Clearly, a calculation of the superfluid density of the disordered $t-J$ model would be the most straightforward way to falsify the spin-charge separation scenario suggested by the above variational study. However, the calculation of the superfluid density of a strongly correlated and disordered system is itself a tough problem. 

In this work we develop a variational theory for the zero temperature superfluid density of the disordered $t-J$ model. This is achieved in two steps. First, we perform an unrestricted variational optimization of an RVB variational ground state for the disordered t-J model. Second, we construct the variational state that describes the paramagnetic current response on such an optimized RVB state. The zero temperature superfluid density of the system is extracted from the second derivative of the variational ground state energy as a function of the external electromagnetic field. We find that $\rho_{s}(0)$ computed in this way is remarkably robust against the disorder effect. More specifically, we find that $\rho_{s}(0)$ is a monotonically increasing function of doping concentration $x$ and is suppressed only gently with the increase of the disorder strength. In fact, we find that the doping and disorder strength dependence of $\rho_{s}(0)$ follows closely that of the total optical weight. This is consistent with the observation in the underdoped cuprates, but is strongly at odd with the behavior in the overdoped cuprates. To be complete, we have also performed a BCS mean field calculation of $\rho_{s}(0)$ in the overdoped regime by ignoring the no double occupancy constraint in the $t-J$ model. Consistent with the result of previous studies adopting a similar weak coupling approach, we find that $\rho_{s}(0)$ is quickly suppressed with the increase of the disorder strength. The strong contrast between the disorder effect in underdoped and the overdoped regime lends strong support to our previous proposal that there exists a Mott transition between a doped-Mott-insulating metal in the underdoped regime and a fermi-liquid-like metal in the overdoped regime around the pseudogap end point. Our numerical result also indicates that the strong electron incoherence induced by the strong correlation effect between the electron acts to protect the d-wave superconductivity in the system from the suppression by the disorder effect.

The paper is organized as follows. In the next section, we introduce the disordered $t-J$ model and describe the variational approach to compute its zero temperature superfluid density. In the third section, we present the BCS mean field theory of $\rho_{s}(0)$ by ignoring the no double occupancy constraint in the $t-J$ model. In the fourth section, we present the numerical results for both the variational theory and the BCS mean field theory. The fifth section is devoted to conclusions and discussions.

\section{The disordered $t-J$ model and its zero temperature superfluid density}
\subsection{The disordered $t-J$ model}
The Hamiltonian of the disordered $t-J$ model studied in this work reads
\begin{eqnarray}
H&=&-t\sum_{\langle i,j \rangle,\sigma}(\hat{c}^{\dagger}_{i,\sigma}\hat{c}_{j,\sigma}+h.c.)\nonumber\\
&-&t'\sum_{\langle\langle i,j \rangle\rangle,\sigma}(\hat{c}^{\dagger}_{i,\sigma}\hat{c}_{j,\sigma}+h.c.)\nonumber\\
&+&J\sum_{\langle i,j \rangle}(\mathbf{S}_{i}\cdot\mathbf{S}_{j}-\frac{1}{4}n_{i}n_{j})+\sum_{i,\sigma}\mu_{i} \hat{c}^{\dagger}_{i,\sigma}\hat{c}_{i,\sigma}
\end{eqnarray} 
 here the electron operator $\hat{c}_{i,\sigma}$ is subjected to the following no double occupancy constraint
\begin{equation}
\sum_{\sigma}\hat{c}^{\dagger}_{i,\sigma}\hat{c}_{i,\sigma}\leq 1
\end{equation}
We use $\sum_{\langle i,j \rangle}$ to denote the sum over the nearest-neighboring(NN) bonds and $\sum_{\langle\langle i,j \rangle\rangle}$ to denote the sum over the next-nearest-neighboring(NNN) bonds. $t$ and $t'$ are the corresponding hopping integrals. $J$ denotes the Heisenberg exchange coupling between NN spins. The last term in the Hamiltonian denotes the disorder potential. In this study, we set $t'/t=-0.3, J/t=0.3$ and use $t$ as the unit of energy. $\mu_{i}$ is assumed to be distributed uniformly in a box region $[-\frac{V}{2},\frac{V}{2}]$.

We assume that the ground state of the disordered $t-J$ model is superconducting. To compute the zero temperature superfluid density of the disordered $t-J$ model, we couple the electron to an electromagnetic(EM) potential through the following Peierls substitution in the kinetic terms
\begin{eqnarray}
H[\mathbf{A}]&=&-\sum_{\langle i,j \rangle,\sigma}(te^{i\mathbf{A}\cdot(\mathbf{r}_{i}-\mathbf{r}_{j})}\hat{c}^{\dagger}_{i,\sigma}\hat{c}_{j,\sigma}+h.c.)\nonumber\\
&-&\sum_{\langle\langle i,j \rangle\rangle,\sigma}(t'e^{i\mathbf{A}\cdot(\mathbf{r}_{i}-\mathbf{r}_{j})}\hat{c}^{\dagger}_{i,\sigma}\hat{c}_{j,\sigma}+h.c.)\nonumber\\
&+&J\sum_{\langle i,j \rangle}(\mathbf{S}_{i}\cdot\mathbf{S}_{j}-\frac{1}{4}n_{i}n_{j})+\sum_{i,\sigma}\mu_{i} \hat{c}^{\dagger}_{i,\sigma}\hat{c}_{i,\sigma}
\end{eqnarray} 
in which $\mathbf{A}$ is a spatial uniform electromagnetic potential and we choose it to be directed in the $x$-direction. The strength of $\mathbf{A}$ is measured in unit of the flux quantum $\Phi_{0}$ as follows
\begin{equation}
A=|\mathbf{A}|=\frac{2\pi}{L}\frac{\Phi}{\Phi_{0}}
\end{equation}
in which $L$ is linear size of the system in the $x$-direction. Note that we have set $e=\hbar=c=a=1$ for convenience, with $a$ denoting the lattice constant of the $CuO_{2}$ plane. The zero temperature superfluid density of the system can be calculated from the ground state energy $E_{g}[A]$ as follows\cite{Scalapino}
\begin{equation}
\rho_{s}(0)=\left. \frac{\partial^{2}E_{g}[A]}{\partial A^{2}} \right|_{L\rightarrow\infty,A=0}
\end{equation}

To compute $\rho_{s}(0)$, we expand $H[\mathbf{A}]$ to the second order in $A$, which reads
\begin{equation}
H[\mathbf{A}]\approx H+H_{1}+H_{2}
\end{equation} 
here
\begin{equation}
H_{1}=-J^{x}_{p}A
\end{equation}
is the paramagnetic coupling and
\begin{eqnarray}
J^{x}_{p}&=&it\sum_{\langle i,j \rangle_{x},\sigma}(\hat{c}^{\dagger}_{i,\sigma}\hat{c}_{j,\sigma}-\hat{c}^{\dagger}_{j,\sigma}\hat{c}_{i,\sigma})\nonumber\\
&+&it'\sum_{\langle\langle i,j \rangle\rangle,\sigma}(\hat{c}^{\dagger}_{i,\sigma}\hat{c}_{j,\sigma}-\hat{c}^{\dagger}_{j,\sigma}\hat{c}_{i,\sigma})
\end{eqnarray} 
is the $x$-component of the paramagnetic current.
\begin{equation}
H_{2}=\frac{1}{2}K^{xx}A^{2}
\end{equation}
is the diamagnetic coupling and
\begin{eqnarray}
K^{xx}&=&t\sum_{\langle i,j \rangle_{x},\sigma}(\hat{c}^{\dagger}_{i,\sigma}\hat{c}_{j,\sigma}+\hat{c}^{\dagger}_{j,\sigma}\hat{c}_{i,\sigma})\nonumber\\
&+&t'\sum_{\langle\langle i,j \rangle\rangle,\sigma}(\hat{c}^{\dagger}_{i,\sigma}\hat{c}_{j,\sigma}+\hat{c}^{\dagger}_{j,\sigma}\hat{c}_{i,\sigma})
\end{eqnarray} 
is the $xx$-component of the inverse effective mass tensor.
 Here $\langle i,j \rangle_{x}$ denotes the nearest neighbor in the $x$-direction. To the first order in $A$, the perturbed ground state of the system is given by
 \begin{equation}
 |\Psi(A)\rangle\approx |0\rangle-A\sum_{n\neq0}\frac{|n\rangle\langle n|J^{x}_{p}| 0\rangle}{E_{0}-E_{n}}
 \end{equation}
 in which $|0\rangle=|\Psi(A=0)\rangle$ denotes the ground state of the system at $A=0$, $|n\rangle$ denotes the corresponding excited state. $E_{0}$ and $E_{n}$ are the eigen energy of these states. It is then straightforward to show that to the second order in $A$
 \begin{equation}
 E_{g}(A)\approx E_{g}(0)+\frac{1}{2}\rho_{s}(0)A^{2}
 \end{equation}
 with
\begin{equation}
\rho_{s}(0)=\langle 0|K^{xx}|0\rangle-2\sum_{n\neq0}\frac{|\langle n|J^{x}_{p}| 0\rangle|^{2}}{E_{n}-E_{0}}
\end{equation} 

Clearly, it is unrealistic to calculate $\rho_{s}(0)$ from Eq.11 directly since it is in general impossible to obtain the full spectrum and the corresponding eigenstates of the disordered $t-J$ model. In the following we develop a variational theory to compute $\rho_{s}(0)$ of the disordered $t-J$ model. In this variational theory, the ground state of the disordered $t-J$ model will be described by a fermionic RVB state. We then propose a variational state that can describe the paramagnetic current response of the system on the optimized RVB state. The zero temperature superfluid density is extracted from the curvature of the variational ground state energy as a function of the external EM field.
 
 \subsection{RVB ground state of the disordered $t-J$ model}
 In this work, we will describe the ground state of the disordered $t-J$ model with the fermionic RVB theory. For this purpose, we represent the constrained electron operator $\hat{c}_{i,\sigma}$ in terms of the charge neutral spinon operator $f_{i,\sigma}$ and the spinless holon operator $b_{i}$ as follows\cite{RVB,PALee}
 \begin{equation}
 \hat{c}_{i,\sigma}=f_{i,\sigma}b^{\dagger}_{i}
 \end{equation}
We note that this is an exact representation of the constrained electron operator if the the spinon and the holon operator satisfy the constraint 
\begin{equation}
\sum_{\sigma}f^{\dagger}_{i,\sigma}f_{i,\sigma}+b^{\dagger}_{i}b_{i}=1
\end{equation}
At the same time, we note that the constrained electron operator is invariant under the following $U(1)$ gauge transformation
\begin{eqnarray}
f_{i,\sigma}&\rightarrow&f_{i,\sigma}e^{i\phi_{i}}\nonumber\\
b_{i}&\rightarrow&b_{i}e^{i\phi_{i}}
\end{eqnarray}
The Hamiltonian of the disordered $t-J$ model in terms of the spinon and the holon operator reads, 
\begin{eqnarray}
H=&-&t\sum_{\langle i,j\rangle,\sigma}(f^{\dagger}_{i,\sigma}f_{j,\sigma}b^{\dagger}_{i}b_{j}+h.c.)\nonumber\\
&-&t'\sum_{\langle\langle i,j\rangle\rangle,\sigma}( f^{\dagger}_{i,\sigma}f_{j,\sigma}b^{\dagger}_{i}b_{j}+h.c.)\nonumber\\
&+&\frac{J}{2}\sum_{\langle i,j \rangle,\sigma,\sigma'}[f^{\dagger}_{i,\sigma}f_{i,\sigma'}f^{\dagger}_{j,\sigma'}f_{j,\sigma}-f^{\dagger}_{i,\sigma}f_{i,\sigma}f^{\dagger}_{j,\sigma'}f_{j,\sigma'}]\nonumber\\
&+&\sum_{i,\sigma}\mu_{i}f^{\dagger}_{i,\sigma}f_{i,\sigma}
\end{eqnarray} 
Introducing the mean field order parameters
\begin{eqnarray}
\chi_{i,j}&=&\langle f^{\dagger}_{i,\uparrow}f_{j,\uparrow}+f^{\dagger}_{i,\downarrow}f_{j,\downarrow}\rangle\nonumber\\
\Delta_{i,j}&=&\langle f^{\dagger}_{i,\uparrow}f^{\dagger}_{j,\downarrow}+f^{\dagger}_{j,\uparrow}f^{\dagger}_{i,\downarrow}\rangle
\end{eqnarray}
and the boson condensate amplitude for the holon
\begin{equation}
\bar{b}_{i}=\langle b_{i} \rangle=\langle b^{\dagger}_{i} \rangle
\end{equation}
the Hamiltonian decouples into its spinon piece 
\begin{eqnarray}
H^{f}_{MF}=&-&\sum_{\langle i,j\rangle,\sigma}(t^{f}_{i,j}f^{\dagger}_{i,\sigma}f_{j,\sigma}+h.c.)+\sum_{i,\sigma}\mu^{f}_{i}f^{\dagger}_{i,\sigma}f_{i,\sigma}\nonumber\\
&-&\sum_{\langle\langle i,j\rangle\rangle,\sigma}(t'^{f}_{i,j} f^{\dagger}_{i,\sigma}f_{j,\sigma}+h.c.)\nonumber\\
&+&\sum_{\langle i,j \rangle}[\Delta^{f}_{i,j}(f^{\dagger}_{i,\uparrow}f^{\dagger}_{j,\downarrow}+f^{\dagger}_{j,\uparrow}f^{\dagger}_{i,\downarrow})+h.c.]
\end{eqnarray}
and the holon piece
\begin{eqnarray}
H^{b}_{MF}=&-&\sum_{\langle i,j\rangle}t^{b}_{i,j}(b^{\dagger}_{i}b_{j}+h.c.)\nonumber\\
&-&\sum_{\langle\langle i,j\rangle\rangle}t'^{b}_{i,j}( b^{\dagger}_{i}b_{j}+h.c.)
\end{eqnarray}
Since the explicit expression of $t^{f}_{i,j}$, $t'^{f}_{i,j}$, $\Delta^{f}_{i,j}$, $t^{b}_{i,j}$ and $t'^{b}_{i,j}$ will not be referenced in the following discussion, we will omit it. 

The RVB variational ground state we adopted to describe the ground state of the disordered $t-J$ model is generated by Gutzwiller projection of the mean field ground state and takes the form of
\begin{equation}
|\mathrm{RVB}\rangle=\mathrm{P}_{\mathrm{G}}| f-\mathrm{BCS}\rangle\otimes |b-\mathrm{Condens}\rangle
\end{equation}
in which $\mathrm{P}_{\mathrm{G}}$ denotes the Gutzwiller projection enforcing the no double occupancy constraint $\sum_{\sigma}f^{\dagger}_{i,\sigma}f_{i,\sigma}+b^{\dagger}_{i}b_{i}=1$. $| f-\mathrm{BCS}\rangle$ is the mean field ground state of $H^{f}_{MF}$, which has the form of the Bogoliubov-de Gennes(BdG) Hamiltonian and can be easily diagonalized. The explicit form of $| f-\mathrm{BCS}\rangle$ reads
\begin{equation}
| f-\mathrm{BCS}\rangle=\exp[\sum_{i,j}a(i,j)f^{\dagger}_{i,\uparrow}f^{\dagger}_{j,\downarrow}] |0\rangle_{f}
\end{equation} 
in which $a(i,j)$ denotes the RVB amplitude of the spinon pair and satisfy the condition
\begin{equation}
a(i,j)=a(j,i)
\end{equation}
as a result of the spin rotational symmetry. $|0\rangle_{f}$ denotes the vacuum state of the spinon. A detailed derivation of the explicit form for $a(i,j)$ from the mean field ansatz $H^{f}_{MF}$ can be found in Appendix A. $|b-\mathrm{Condens}\rangle$ denotes the holon condensate and takes the form of 
\begin{equation}
|b-\mathrm{Condens}\rangle=\left(\sum_{i}\bar{b}_{i}b^{\dagger}_{i}\right)^{N_{b}}|0\rangle_{b}
\end{equation}
in which $N_{b}$ is the number of doped holes, $|0\rangle_{b}$ denotes the vacuum of the holon Hilbert space. We note that the U(1) gauge redundancy allow us to choose a gauge so that the Bose condensate amplitude $\bar{b}_{i}$ is positive definite.

In principle, we can take the parameters in $H^{f}_{MF}$ and $H^{b}_{MF}$, namely $t^{f}_{i,j}$, $t'^{f}_{i,j}$, $\Delta^{f}_{i,j}$, $t^{b}_{i,j}$ and $t'^{b}_{i,j}$, as the variational parameters to be optimized in $|\mathrm{RVB}\rangle$. Here instead we will take the RVB amplitude $a(i,j)$ and the bose condensate amplitude $\bar{b}_{i}$ as variational parameter directly. The advantage for such a generalization is twofold. First, the RVB state with arbitrary value of $a(i,j)$ and $\bar{b}_{i}$ represents the most general RVB state that is consistent with the spin rotational symmetry of the system. In other words, the hopping and the pairing parameters in the mean field ansatz are now not restricted to the NN and the NNN bonds, but can be arbitrarily long ranged. This can improve the descriptive power of the variational state. Second, the optimization of the RVB amplitude and bose condensate amplitude is simpler than that of the parameters involved in the mean field ansatz. In the latter case, the gradient of the ground state wave function over the variational parameters can only be computed through the first order perturbation expansion of the mean field Hamiltonian, which is numerically expensive and storage demanding. On the other hand, the computation of the gradient becomes almost trivial when we choose $a(i,j)$ and $\bar{b}_{i}$ as the variational parameter.

One disadvantage of choosing $a(i,j)$ and $\bar{b}_{i}$ as the variational parameter is that the number of variational parameters to be optimized is now dramatically increased. More specifically, on a finite cluster with $N$ sites, there will be $\frac{N(N-1)}{2}+N$ real parameters to be optimized. Among these there are $\frac{N(N-1)}{2}$ parameters coming from the RVB amplitude $a(i,j)$ and $N$ parameters coming from the bose condensate amplitude $\bar{b}_{i}$. We note both the RVB amplitude $a(i,j)$ and the bose condensate amplitude $\bar{b}_{i}$ are assumed to be real number. This is reasonable since we expect the optimized variational ground state to be time reversal invariant. With the advance of variational optimization technique, the unrestricted optimization of $N^{2}$ parameters becomes accessible even on rather large clusters with the finite-depth BFGS algorithm\cite{Yang,Li,Li2,Rong1,Rong2}.

To carry out the variational optimization on $|\mathrm{RVB}\rangle$, we expand it in a orthornomal basis $|R\rangle$. Here we use the following Ising basis
\begin{equation}
|R\rangle=\prod^{N_{e}}_{k=1}f^{\dagger}_{i_{k},\uparrow}f^{\dagger}_{j_{k},\downarrow}|0\rangle_{f}\otimes \prod^{N_{b}}_{k'=1}b^{\dagger}_{l_{k'}}|0\rangle_{b}
\end{equation}
in which $i_{k}$ denotes the position of the $k$-th up-spin electrons, $j_{k}$ denotes the position of the $k$-th down-spin electrons, $l_{k'}$ denotes the position of the $k'$-th unoccupied sites. Note that we have equal number of up-spin and down-spin electron($N_{e}$) in the ground state. The wave function amplitude of $|\mathrm{RVB}\rangle$ in this basis is given by
\begin{equation}
\Psi(R)=\mathrm{Det}[\mathbf{a}]\times \prod_{k=1}^{N_{b}} \bar{b}_{l_{k}}
\end{equation}
in which $\mathbf{a}$ is a $N_{e}\times N_{e}$ matrix of the form
\begin{eqnarray}
\mathbf{a}=\left(\begin{array}{cccc}a(i_{1},j_{1}) & . & . & a(i_{1},j_{N_{e}}) \\. & . & . & . \\ .& . & . & . \\   . & . & . & .\\
a(i_{N_{e}},j_{1}) & . & . &a(i_{N_{e}},j_{N_{e}})\nonumber
 \end{array}\right)
\end{eqnarray}
The variational ground state energy is then given by
\begin{equation}
E=\langle H \rangle_{\Psi}=\frac{\langle\Psi| H |\Psi\rangle}{\langle \Psi |\Psi \rangle}=\frac{\sum_{R}|\Psi(R)|^2 E_{loc}(R)}{\sum_{R}|\Psi(R)|^2}
\end{equation}
in which the local energy $E_{loc}(R)$ is defined as
\begin{equation}
E_{loc}(R)=\sum_{R'}\langle R |H| R' \rangle \frac{\Psi(R')}{\Psi(R)}
\end{equation}
The gradient of the variational energy with respect to the variational parameters is given by
\begin{equation}
\nabla E=2[ \langle \nabla \ln \Psi(R) E_{loc}(R) \rangle_{\Psi}-E\langle \nabla \ln \Psi(R) \rangle_{\Psi}]
\end{equation}
Both $E$ and $\nabla E$ can be computed by the standard Monte Carlo sampling over the distribution generated by $|\Psi(R)|^2$. In the calculation of the energy gradient, the key quantity to be computed is $\nabla \ln \Psi(R)$. For the gradient on the RVB amplitude, we have
\begin{equation}
\nabla\ln \Psi(R)=\mathbf{Tr} [\nabla \mathbf{a} \mathbf{a}^{-1}]
\end{equation}
The matrix elements of $\nabla \mathbf{a}$ is trivial and is either 1 or 0. The gradient of $\Psi(R)$ on the bose condensate amplitude is given by
\begin{equation}
\frac{\partial}{\partial \bar{b}_{i}}\ln \Psi(R)=\frac{1}{\bar{b}_{i}}\sum^{N_{b}}_{k=1}\delta_{l_{k},i}
\end{equation}

In the finite-depth BFGS algorithm\cite{Li,Rong2}, an iterative approximation for the inverse of the Hessian matrix is generated from the gradient of variational energy as follows
\begin{equation}
\mathbf{B}_{k+1}=\left(\mathbf{I}-\frac{\mathbf{s}_{k}\mathbf{y}^{T}_{k}}{\mathbf{y}^{T}_{k}\mathbf{s}_{k} } \right)\mathbf{B}_{k}\left(\mathbf{I}-\frac{\mathbf{y}_{k}\mathbf{s}^{T}_{k}}{\mathbf{y}^{T}_{k}\mathbf{s}_{k} } \right)+\frac{\mathbf{s}_{k}\mathbf{s}^{T}_{k}}{\mathbf{y}^{T}_{k}\mathbf{s}_{k} }
\end{equation}
here $k=0,1,2.....$ is the iteration step index,
\begin{eqnarray}
\mathbf{s}_{k}&=&\bm{\alpha}_{k+1}-\bm{\alpha}_{k}\nonumber\\
\mathbf{y}_{k}&=&\nabla E_{k+1}-\nabla E_{k}
\end{eqnarray}  
are the difference between successive variational parameters and the difference between successive energy gradients at iteration step $k$. Here $\bm{\alpha}_{k=0}$ is the initial guess for the variational parameters. $\nabla E_{k=0}$ is the energy gradient calculated at the starting point. The inverse Hessian matrix $\mathbf{B}_{k}$ is initially set to be the identity matrix. Using such an iterative approximation on the inverse of the Hessian matrix, the variational parameters are updated as follows
\begin{equation}
\bm{\alpha}_{k+1}=\bm{\alpha}_{k}+\delta\ \mathbf{B}_{k} \nabla E_{k}
\end{equation}  
in which $\delta$ is the step length chosen by trial and error. In practice, we restart the BFGS iteration every $K$ step. Since all we need in the BFGS iteration is the product of the matrix $\mathbf{B}_{k}$ with the vector $\nabla E_{k}$ or $\mathbf{y}_{k}$ and the inner product between the vector $\mathbf{s}_{k}$ and $\mathbf{y}_{k}$, we only need to store the $2K$ vectors $\mathbf{s}_{k}$ and $\mathbf{y}_{k}$ in the $K$-depth BFGS algorithm. The needed matrix-vector and vector inner product can be computed recursively using these $2K$ vectors. 

 \subsection{Variational description of the paramagnetic current response on $|\mathrm{RVB}\rangle$}
Now we develop a variational theory for the paramagnetic current response on $|\mathrm{RVB}\rangle$. The basic strategy is to propose a variational subspace that can saturate the optical weight of the paramagnetic current response from $|\mathrm{RVB}\rangle$. The simplest variational subspace that saturates the optical weight of the paramagnetic current response contains only two state vectors, namely $|\mathrm{RVB}\rangle$ and $H_{1}|\mathrm{RVB}\rangle$. The variational ground state in the presence of the EM field can be expressed with these two state vectors as 
\begin{eqnarray}
|\Psi(A)\rangle&=&|\mathrm{RVB}\rangle+\lambda(A)H_{1}|\mathrm{RVB}\rangle
\end{eqnarray}
in which $\lambda(A)$ is a $A$-dependent variational parameter to be determined by the optimization of the variational ground state energy, which is given by
\begin{equation}
E_{g}(A)=\frac{\langle\Psi(A)|H[\mathbf{A}]|\Psi(A)\rangle}{\langle\Psi(A)|\Psi(A)\rangle}
\end{equation} 
$\rho_{s}(0)$ can then be extracted from the second derivative of $E_{g}(A)$ over $A$. It is easy to show that the optimization of $E_{g}(A)$ reduces to the solution of the following generalized eigenvalue problem
\begin{equation}
\left(\begin{array}{cc}H_{1,1} & H_{1,2} \\H_{2,1} & H_{2,2}\end{array}\right)\left(\begin{array}{c}\varphi_{1} \\\varphi_{2}\end{array}\right)=E\left(\begin{array}{cc}O_{1,1} & O_{1,2} \\O_{2,1} & O_{2,2}\end{array}\right)\left(\begin{array}{c}\varphi_{1} \\\varphi_{2}\end{array}\right)
\end{equation}
Here we introduce the abbreviation $|1\rangle=|\mathrm{RVB}\rangle$ and $|2\rangle=H_{1}|\mathrm{RVB}\rangle$. 
\begin{equation}
H_{i,j}=\langle i|H[\mathbf{A}]|j\rangle, \ \ \ \ O_{i,j}=\langle i|j\rangle
\end{equation}
are the Hamiltonian matrix element and the overlap matrix element between these two basis vectors, which are in general non-orthonormal. These matrix elements can be computed from Monte Carlo sampling on $|1\rangle$ and $|2\rangle$. A detailed discussion on this point will be given below.

In the presence of the disorder potential, the translational and the rotational symmetry of the system will be explicitly broken. As a result, the system will rearrange its paramagnetic current response on each bond of the lattice. The variational form Eq.36 is then too restrictive to address such a rearrangement. To propose a spatially nonuniform generalization of Eq.36, we expand $H_{1}|\mathrm{RVB}\rangle$ as follows
\begin{eqnarray}
&&H_{1}|\mathrm{RVB}\rangle=-AJ^{x}_{p}|\mathrm{RVB}\rangle\nonumber\\
&&\ \ \ \ \ \ \ \  \  =-iAt\sum_{\langle i,j\rangle_{x},\sigma}(\hat{c}^{\dagger}_{i,\sigma}\hat{c}_{j,\sigma}-\hat{c}^{\dagger}_{j,\sigma}\hat{c}_{i,\sigma})|\mathrm{RVB}\rangle\nonumber\\
&&\ \ \ \ \ \ \ \ \ \ \  -iAt'\sum_{\langle\langle i,j\rangle\rangle,\sigma}(\hat{c}^{\dagger}_{i,\sigma}\hat{c}_{j,\sigma}-\hat{c}^{\dagger}_{j,\sigma}\hat{c}_{i,\sigma})|\mathrm{RVB}\rangle\nonumber\\
&&\ \ \ \ \ \ \ \ \ \  \ =\sum^{8N}_{m=1}\varphi^{(0)}_{m}\ \left[\sum_{\sigma}\hat{c}^{\dagger}_{i_{m},\sigma}\hat{c}_{j_{m},\sigma}\right]|\mathrm{RVB}\rangle
\end{eqnarray}  
$\sum_{m}$ denotes the sum over all the NN and NNN bonds on the lattice, with $m$ denoting the bond index. We note that the bond here is directed and there are in total $8N$ such bonds on a finite cluster with $N$ sites. $i_{m}$ and $j_{m}$ denote the starting and the end point of the $m$-th bond. The expansion coefficient $\varphi^{(0)}_{m}$ is irrelevant for the following discussion and will be omitted. Clearly, in the presence of the disorder potential $\varphi^{(0)}_{m}$ will be modified. With such a consideration in mind, we propose the following spatially nonuniform generalization of Eq.36
\begin{eqnarray}
|\Psi(A)\rangle&=&|\mathrm{RVB}\rangle+\sum^{8N}_{m=1}\varphi_{m}(A) \left[\sum_{\sigma}\hat{c}^{\dagger}_{i_{m},\sigma}\hat{c}_{j_{m},\sigma}\right]|\mathrm{RVB}\rangle\nonumber\\
\end{eqnarray} 
in which the $A$-dependent coefficient $\varphi_{m}(A)$ is to be determined by optimization of the variational energy $E_{g}(A)$ as defined by Eq.37. This is again equivalent to the solution of a generalized eigenvalue problem, which takes the form of
\begin{equation}
\sum^{8N}_{m'=0}H_{m,m'}\varphi_{m'}=E\sum^{8N}_{m'=0}O_{m,m'}\varphi_{m'}
\end{equation}
Here we introduce the abbreviation 
\begin{equation}
|m\rangle=\left\{\begin{aligned}
|\mathrm{RVB}\rangle& \ \ \ \ m=0\\
\left[\sum_{\sigma}\hat{c}^{\dagger}_{i_{m},\sigma}\hat{c}_{j_{m},\sigma}\right]|\mathrm{RVB}\rangle& \ \ \ \ m\neq0
\end{aligned}\right.
\end{equation} 
The matrix elements $H_{m,m'}$ and $O_{m,m'}$ are defined as
\begin{equation}
H_{m,m'}=\langle m|H[\mathbf{A}]|m'\rangle, \ \ \ \ O_{m,m'}=\langle m|m'\rangle
\end{equation}
These matrix elements can be computed efficiently with the re-weighting method\cite{Li1}. More specifically, denoting the wave function amplitude of the basis vector $|m\rangle$ in the Ising basis $|R\rangle$ as $\psi_{m}(R)$, we have 
\begin{eqnarray}
&&\langle m|H[\mathbf{A}]|m'\rangle=S\frac{\langle m|H[\mathbf{A}]|m'\rangle}{\sum_{m}\langle m |m \rangle}\nonumber\\
&&\ \ \ \ \ \ =S\frac{\sum_{R,R'}\psi^{*}_{m}(R)\langle R|H[\mathbf{A}]|R'\rangle\psi_{m'}(R')}{\sum_{R}W(R)}\nonumber\\
&&\ \ \ \ \ \ =S\frac{\sum_{R}W(R)\sum_{R'}\frac{\psi^{*}_{m}(R)\langle R|H[\mathbf{A}]|R'\rangle\psi_{m'}(R')}{W(R)}}{\sum_{R}W(R)}\nonumber\\
&&\ \ \ \ \ \ =S\frac{\sum_{R}W(R)\frac{v^{*}_{m}(R)E_{loc,m'}(R)}{\sum_{m}|v_{m}(R)|^{2}}}{\sum_{R}W(R)}
\end{eqnarray} 
Here $S=\sum_{m}\langle m|m\rangle$, $W(R)=\sum_{m}|\psi_{m}(R)|^{2}$ is the joint distribution of the state vectors $|m\rangle$ in the variational subspace,
\begin{equation}
v_{m}(R)=\frac{\psi_{m}(R)}{\psi_{0}(R)}
\end{equation}
is the $m$-component of a length-$8N+1$ vector. $\psi_{0}(R)$ is the wave function amplitude of the reference state, which is chosen here to be $|0\rangle=|\mathrm{RVB}\rangle$, $E_{loc,m'}(R)$ is the local energy vector. It is given by
\begin{equation}
E_{loc,m'}(R)=\sum_{R'}\langle R|H[\mathbf{A}]|R'\rangle\frac{\psi_{0}(R')}{\psi_{0}(R)}v_{m'}(R')
\end{equation}  
Similarly, we have
\begin{eqnarray}
&&\langle m|m'\rangle=S\frac{\langle m|m'\rangle}{\sum_{m}\langle m |m \rangle}\nonumber\\
&&\ \ \ \ \ =S\frac{\sum_{R}W(R)\frac{v^{*}_{m}(R)v_{m'}(R)}{\sum_{m}|v_{m}(R)|^{2}}}{\sum_{R}W(R)}
\end{eqnarray} 
Thus, the computation of $H_{m,m'}$ and $O_{m,m'}$ becomes highly parallelized with the introduction of the vector $v_{m}(R)$ and can be completed in a single Monte Carlo sampling over the joint distribution $W(R)$. The same vector $v_{m}(R)$ is also used in the Monte Carlo sampling of $W(R)$, for which we have
\begin{equation}
\frac{W(R')}{W(R)}=\left|\frac{\psi_{0}(R')}{\psi_{0}(R)}\right|^{2}\frac{\sum_{m}|v_{m}(R')|^{2}}{\sum_{m} |v_{m}(R)|^{2}}
\end{equation} 

The variational ground state energy of the system in the electromagnetic field, namely $E_{g}(A)$, is given by the lowest eigenvalue of the generalized eigenvalue problem Eq.42. With $E_{g}(A)$ we can then compute the zero temperature superfluid density $\rho_{s}(0)$ with Eq.5. In addition to $\rho_{s}(0)$, we can also compute the total optical weight by calculating $E_{g}(A)$ with $|\Psi(A)$ fixed at $|\Psi(A=0)\rangle$, namely totally suppressing the paramagnetic current response. We note that $|\Psi(A)\rangle$ contains quantum fluctuation correction from the kinetic energy term and is thus different from $|\mathrm{RVB}\rangle$ even if we set $A=0$. In the following we denote the total optical weight computed this way as $K$.

\section{The zero temperature superfluid density of a disordered d-wave BCS superconductor}
To illustrate the role of the strong correlation effect, we now present a mean field theory of the disordered $t-J$ model. The no double occupancy constraint on the electron operator is totally ignored in such a mean field treatment. This is nothing but the extensively studied Bogliubov-de Gennes(BdG) theory of a dirty d-wave BCS superconductor, which is thought to be relevant for the overdoped cuprates\cite{Alloul,Xiang,Lizx,Kivelson}. It is well known that the d-wave pairing is fragile against impurity scattering as a result of its non-trivial phase structure. 

The zero temperature superfluid density in the BdG theory can be computed with the perturbation theory. For this purpose, we first decouple the exchange term of the $t-J$ model as follows
\begin{equation}
\mathbf{S}_{i}\cdot\mathbf{S}_{j}-\frac{1}{4}n_{i}n_{j}\approx-[\Delta_{i,j}(c^{\dagger}_{i,\uparrow}c^{\dagger}_{j,\downarrow}+c^{\dagger}_{j,\uparrow}c^{\dagger}_{i,\downarrow})+h.c.]+C
\end{equation} 
in which $C$ is an unimportant constant and
\begin{equation}
\Delta_{i,j}=\langle c_{j,\downarrow}c_{i,\uparrow}\rangle
\end{equation}   
Note that here we have removed the hat on the electron operator $c_{i,\sigma}$, which is now a free fermion operator.   
Inserting this mean field decoupling into the $t-J$ Hamiltonian we have
\begin{eqnarray}
H_{BdG}&=&-t\sum_{\langle i,j \rangle,\sigma}(c^{\dagger}_{i,\sigma}c_{j,\sigma}+h.c.)+\sum_{i,\sigma}\mu_{i} c^{\dagger}_{i,\sigma}c_{i,\sigma}\nonumber\\
&-&t'\sum_{\langle\langle i,j \rangle\rangle,\sigma}(c^{\dagger}_{i,\sigma}c_{j,\sigma}+h.c.)\nonumber\\
&-&J\sum_{\langle i,j \rangle}[\Delta_{i,j}(c^{\dagger}_{i,\uparrow}c^{\dagger}_{j,\downarrow}+c^{\dagger}_{j,\uparrow}c^{\dagger}_{i,\downarrow})+h.c.]
\end{eqnarray}
which can be put into the following matrix form
\begin{equation}
H_{BdG}=\bm{\psi}^{\dagger}\mathbf{M}\bm{\psi}
\end{equation}
in which 
\begin{equation}
\bm{\psi}^{\dagger}=(c^{\dagger}_{1,\uparrow},...,c^{\dagger}_{N,\uparrow},c_{1,\downarrow},....,c_{N,\downarrow})
\end{equation}
$\mathbf{M}$ is a $2N\times 2N$ Hermitian matrix whose matrix element is given by the hopping and pairing parameters appearing in Eq.52.
$H_{BdG}$ can be diagonalized by the following unitary transformation
\begin{equation}
\bm{\psi}=\mathbf{U}\bm{\gamma}
\end{equation} 
in which $\mathbf{U}$ is a $2N\times2N$ unitary matrix,
\begin{equation}
\bm{\gamma}=(\gamma_{1},...,\gamma_{N},\gamma_{N+1},....,\gamma_{2N})^{T}
\end{equation}
The diagonalized Hamiltonian reads
\begin{equation}
H_{BdG}=\sum_{n=1}^{2N}\epsilon_{n}\gamma^{\dagger}_{n}\gamma_{n}
\end{equation}
in which $\epsilon_{n}$ is the $n$-th eigenvalue of matrix $\mathbf{M}$. We note that as a result of the spin rotational symmetry, these eigenvalues always appear in $\pm$ pairs. Here we assume that the first $N$ eigenvalues are negative. The self-consistent equation for the pairing potential is given by
\begin{equation}
\Delta_{i,j}=\sum_{n=1}^{N}U^{*}_{j+N,n}U_{i,n}
\end{equation}
Here $U_{i,n}$ is the matrix element of the unitary matrix $\mathbf{U}$. This equation can be solved numerically by iteration. To maintain a comparble size in the pairing gap as that predicted by the RVB theory, in our mean field treatment we will increase the strength of the Heisenberg exchange coupling to a value of $J=t$.

With the eigenvalues and the eigenvectors of $H_{BdG}$ at hand, we can compute the zero temperature superfluid density with the linear response theory. The Kubo formula for $\rho_{s}(0)$ reads
\begin{equation}
\rho_{s}(0)=\langle 0| K^{xx} |0\rangle-2\left.\sum_{n\neq 0}\frac{|\langle n| j^{x}_{\mathbf{q}} |0\rangle|^{2}}{E_{n}-E_{0}}\right|_{|\mathbf{q}|\rightarrow 0}
\end{equation}
in which $|0\rangle$ and $|n\rangle$ denote the ground and the excited state the of BdG Hamiltonian. 
\begin{equation}
K^{xx}=\sum_{\mathbf{k},\sigma}\frac{\partial^{2}\epsilon_{\mathbf{k}}}{\partial k^{2}_{x}}c^{\dagger}_{\mathbf{k},\sigma}c_{\mathbf{k},\sigma}
\end{equation}
is the $xx$-component of the inverse effective mass tensor. 
\begin{equation}
\epsilon_{\mathbf{k}}=-2t(\cos k_{x}+\cos k_{y})-4t'\cos k_{x}\cos k_{y}
\end{equation}
is the bare band dispersion. $j^{x}_{\mathbf{q}} $ is the $x$-component of the paramagnetic current at momentum $\mathbf{q}$. It is given by
\begin{equation}
j^{x}_{\mathbf{q}}=\sum_{\mathbf{k},\sigma}\left.\frac{\partial \epsilon_{\mathbf{k}}}{\partial k_{x}}\right|_{\mathbf{k}+\frac{\mathbf{q}}{2}}c^{\dagger}_{\mathbf{k+q},\sigma}c_{\mathbf{k},\sigma} 
\end{equation}

 We note that a small but nonzero momentum $\mathbf{q}$ must be introduced to compute the paramagnetic current response of a BCS superconductor, as a result of the fact that the electron momentum is strictly conserved in the clean limit in the BCS mean field approximation. The Kubo formula Eq.59 for $\rho_{s}(0)$ is only valid only when we work at a finite $\mathbf{q}$. This is very similar to the situation when we want to compute the uniform spin susceptibility of a spin rotational invariant system, as we also need to introduce a small nonzero wave vector. Otherwise we have to adopt the much more awkward canonical ensemble with conserved magnetization in the computation.  

 \section{Numerical results}
 \subsection{Model parameter specification}
Our variational computation of $\rho_{s}(0)$ is performed on a $L\times L$ square lattice cluster with periodic boundary condition in the $x$-direction and anti-periodic boundary condition in the $y$-direction. The majority part of the numerical data is obtained on a $L=16$ cluster. We find that the finite size effect on the computed $\rho_{s}(0)$ is already small on cluster of such a size. The smallness of the finite size effect in the computed $\rho_{s}(0)$ can be attributed to two reasons. First, unlike the non-interacting case, the electron in the $t-J$ model is subjected to strong incoherence effect as a result of the strong correlation effect. There is thus no bottleneck for the relaxation of electron momentum. In fact, our calculation shows that a substantial part of the total optical weight remains incoherent in the ground state even in the absence of disorder potential. This is very different from the conventional BCS superconductor, in which all Drude weight should condense at zero temperature in the clean limit. Second, the disorder potential provides additional reservior for the relaxation of electron momentum. In other words, the disorder potential cutoffs the coherent propagation of the electron in the system and thus acts to reduce the finite size effect.   
 
In our computation, the disorder strength $V/t$ is varied in the range of $V/t\in[0,2]$. We note that the disorder potential with $V=2t$ is already strong enough to suppress the d-wave superconductivity almost totally if we ignore the no double occupancy constraint in the $t-J$ model. We also note that the average over the disorder realization is not essential for the computed $\rho_{s}(0)$. This again can be attributed to the strong intrinsic electron incoherence in the $t-J$ model. We thus assume the disorder potential to take the following form  
\begin{equation}
\mu_{i}=V(r_{i}-0.5)
\end{equation}
in which $r_{i}$ is random number uniformly distributed in the range of $[0,1]$. We will fix $r_{i}$ and tune the disorder strength by varying $V$. We find that different realizations of the disorder potential with the same value for $V$ produce essentially the same result for $\rho_{s}(0)$. 

Unlike the strongly correlated $t-J$ model, the finite size effect in the BCS mean field theory is very large. This is especially so when the disorder potential is weak and that the pairing gap has node. In our calculation, the $|\mathbf{q}|\rightarrow 0$ limit in Eq.59 is taken by setting $\mathbf{q}$ to be the smallest possible momentum in the Brillouin zone. To reduce the finite size effect without diagonalizing huge matrix, we adopt the supercell trick by replicating the disorder configuration in the relative small supercell of $N$ sites periodically to a large number of(for example $N_{s}$) supercells. We then Fourier transform the Hamiltonian matrix of the system, which is of rank $2NN_{s}$ and bring it into block diagonal form with each block of rank $2N$. The details of the supercell trick can be found in the literature\cite{Ghosal}. In our computation, we have set $N=L\times L=1024$ and $N_{s}=s\times s=200$, resulting in a huge lattice containing as many as $6400\times6400$ sites. We find that such a huge lattice is required to recover the intrinsic paramagnetic current response of the d-wave BCS superconductor, which scales super-linearly with the momentum $\mathbf{q}$ in the long wavelength limit\cite{Ghosal}. 

\subsection{The optimized RVB state at zero field and its off-diagonal-long-range-order}
The unrestricted optimization of the RVB state for the disordered $t-J$ model has already been attempted in our previous study, in which we have adopted a short-ranged mean field ansatz of the form Eq.20. Here we use such an optimized short-ranged mean field ansatz to generate the initial guess for the RVB amplitude $a(i,j)$ of the generalized RVB state. We find that the property of the optimized generalized RVB state is essentially unchanged from that of the optimized RVB state with a short-ranged mean field ansatz. In our previous study, it is found that the d-wave phase structure of the spinon pairing potential is strictly preserved both locally and globally even in the presence of rather strong disorder potential. On the other hand, the amplitude of the spinon pairing potential does exhibit significant spatial modulation. We find that such modulation can be understood as the secondary effect of the spatial modulation in the local holon density, which experiences the disorder potential more directly. In our generalized RVB state, it is impossible to define the spinon pairing potential, which is only a theoretical construction. Here we will present the result for the physically more relevant quantity, namely the off-diagonal-long-range-order(ODLRO) in the optimized RVB state.

The ODLRO for the d-wave superconducting state of the disordered $t-J$ model is defined as follows
\begin{equation}
F=\sqrt{\frac{1}{N}\sum_{i}\langle \hat{\Delta}_{i+\mathbf{R}}^{\dagger}\hat{\Delta}_{i}\rangle}
\end{equation} 
in which $\mathbf{R}$ denotes the largest distance on the $L\times L$ cluster under the periodic boundary condition, $\hat{\Delta}_{i}$ is the pairing field centered on site $i$. It is defined as
\begin{equation}
\hat{\Delta}_{i}=\frac{1}{4}[\hat{\Delta}_{i,i+x}+\hat{\Delta}_{i,i-x}-\hat{\Delta}_{i,i+y}-\hat{\Delta}_{i,i-y}]
\end{equation}
in which
\begin{equation}
\hat{\Delta}_{i,j}=\hat{c}_{i,\uparrow}\hat{c}_{j,\downarrow}+\hat{c}_{j,\uparrow}\hat{c}_{i,\downarrow}
\end{equation}
is the paring field on the bond connecting site $i$ and $j$.
 
The evolution of $F$ with disorder strength is shown in Fig.1 for $x=0.236$. The ODLRO predicted by the generalized RVB ansatz is slightly larger than that predicted by RVB state with a short-ranged mean field ansatz. More specifically, the ODLRO is found to decrease monotonically with $V/t$. However, the level of the reduction is rather small, amounting to only about 13 percent of its clean limit value at the strongest disorder potential of $V/t=2$. Such a disorder potential is already about than 6 times stronger than the bare Heisenberg exchange coupling $J$. In the plain B-dG treatment, the d-wave superconductivity is almost totally suppressed by such a strong disorder potential. Thus the electron correlation effect is a strong protector of the d-wave superconducting order in the $t-J$ model. 

\begin{figure}
\includegraphics[width=8cm]{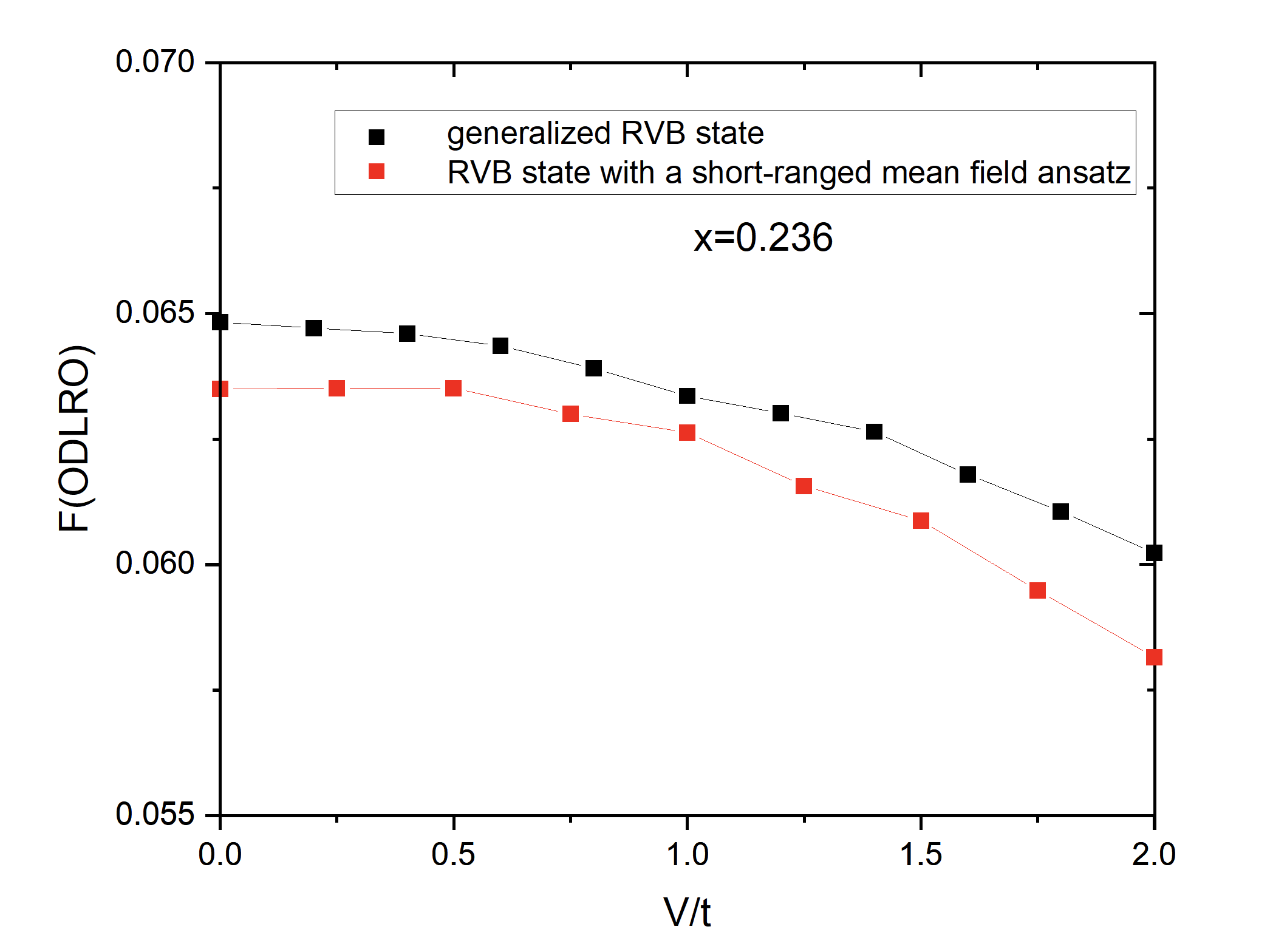}
\caption{The evolution of the ODLRO with the strength of the disorder potential at $x=0.236$ in the optimized RVB ground state. The calculation is done on a $12\times12$ lattice with periodic boundary condition in $x$-direction and anti-periodic boundary condition in $y$-direction. For purpose of comparison, we have also presented our previous result obtained on the RVB state with a short-ranged mean field ansatz.}
\end{figure}  

To have a complete picture of the disorder effect in the 2D $t-J$ model, we have mapped out the whole doping-disorder strength evolution of the ODLRO in Fig.2. The ODLRO is found to decrease only gently with the increase of the disorder potential at all doping concentration. More specifically, we find that the reduction in the ODLRO never exceed 20 percent of its clean limit value even at a disorder strength that is more than 6 times stronger than the Heisenberg exchange coupling $J$. This remarkable robustness of the d-wave pairing is thus a genuine characteristic of the 2D $t-J$ model and is surely beyond the BCS theory description.  

\begin{figure}
\includegraphics[width=8cm]{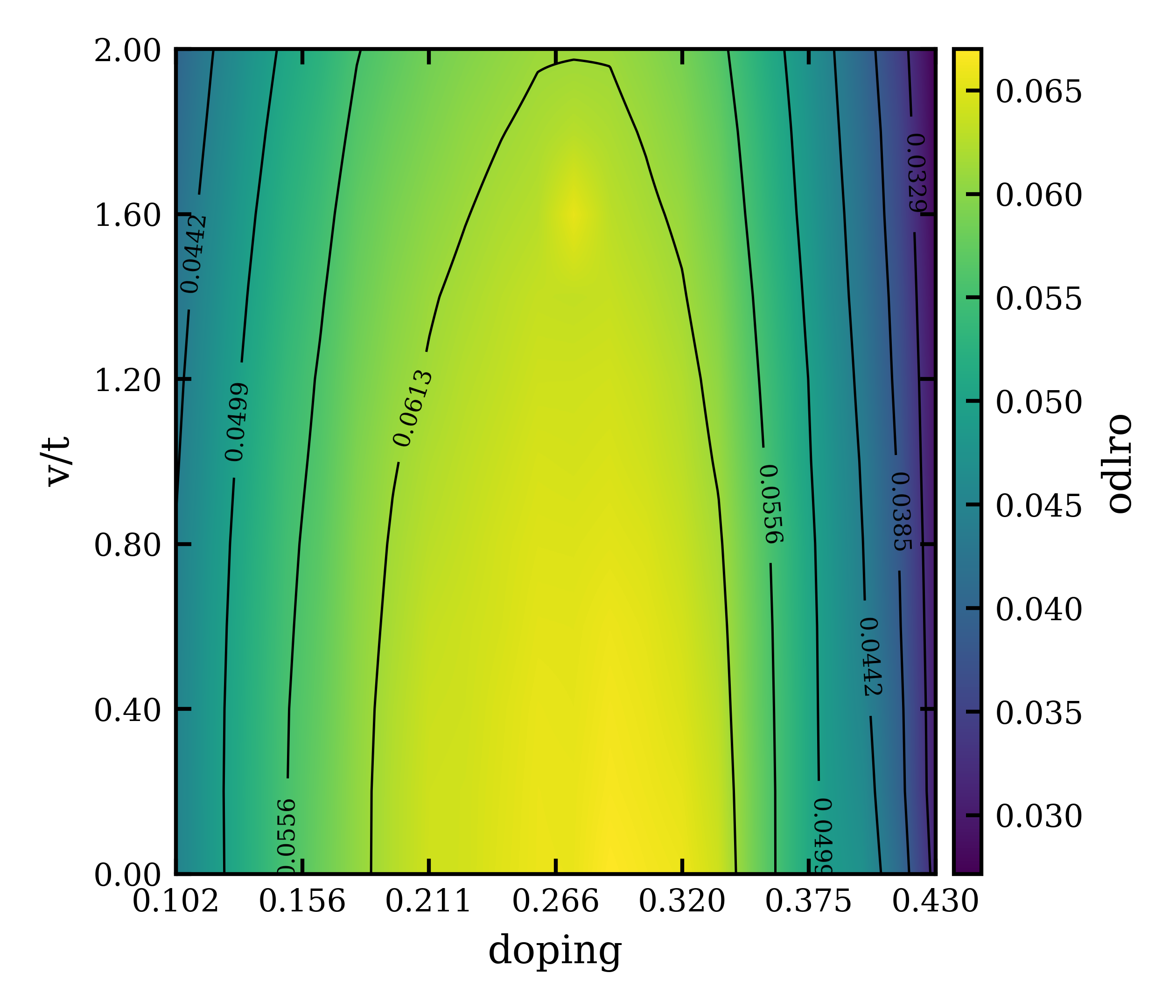}
\caption{The doping-disorder strength evolution of the ODLRO computed in the optimized RVB ground state of the 2D $t-J$ model. The calculation is done on a $16\times16$ lattice with periodic boundary condition in the $x$-direction and anti-periodic boundary condition in the $y$-direction.}
\end{figure}  

\subsection{$\rho_{s}(0)$ of the optimized RVB state and the ratio for the condensed spectral weight}
We have used both the two component wave function Eq.36 and its generalization Eq.41 to compute $\rho_{s}(0)$. The evolution of $\rho_{s}(0)$ with doping and disorder strength computed from both variational schemes are shown in Fig.3. While $\rho_{s}(0)$ computed from both variational schemes differ quantitatively, they exhibit almost identical trend in their doping and disorder strength dependence. More specifically, $\rho_{s}(0)$ is found to increase monotonically with doping concentration at any given disorder strength. At the same time, $\rho_{s}(0)$ is only gently suppressed with the increase of the disorder strength. In particular, the reduction in $\rho_{s}(0)$ never exceed 20 percent of its clean limit value even at the strongest disorder potential of $V=2t$ that we have attempted, whence the d-wave BCS pairing would be almost totally suppressed if we ignore the no double occupancy constraint in the $t-J$ model.

\begin{figure}
\includegraphics[width=8cm]{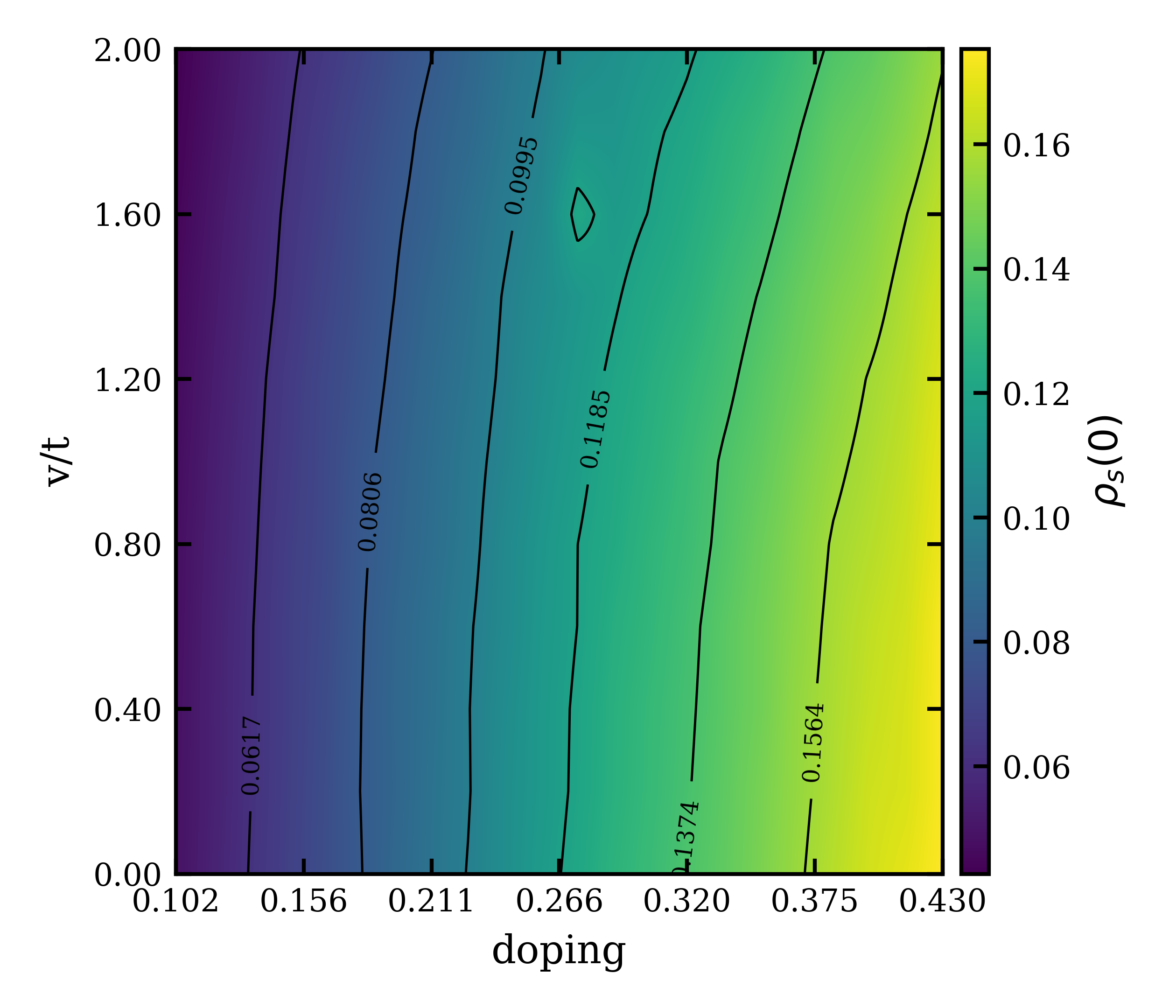}
\includegraphics[width=8cm]{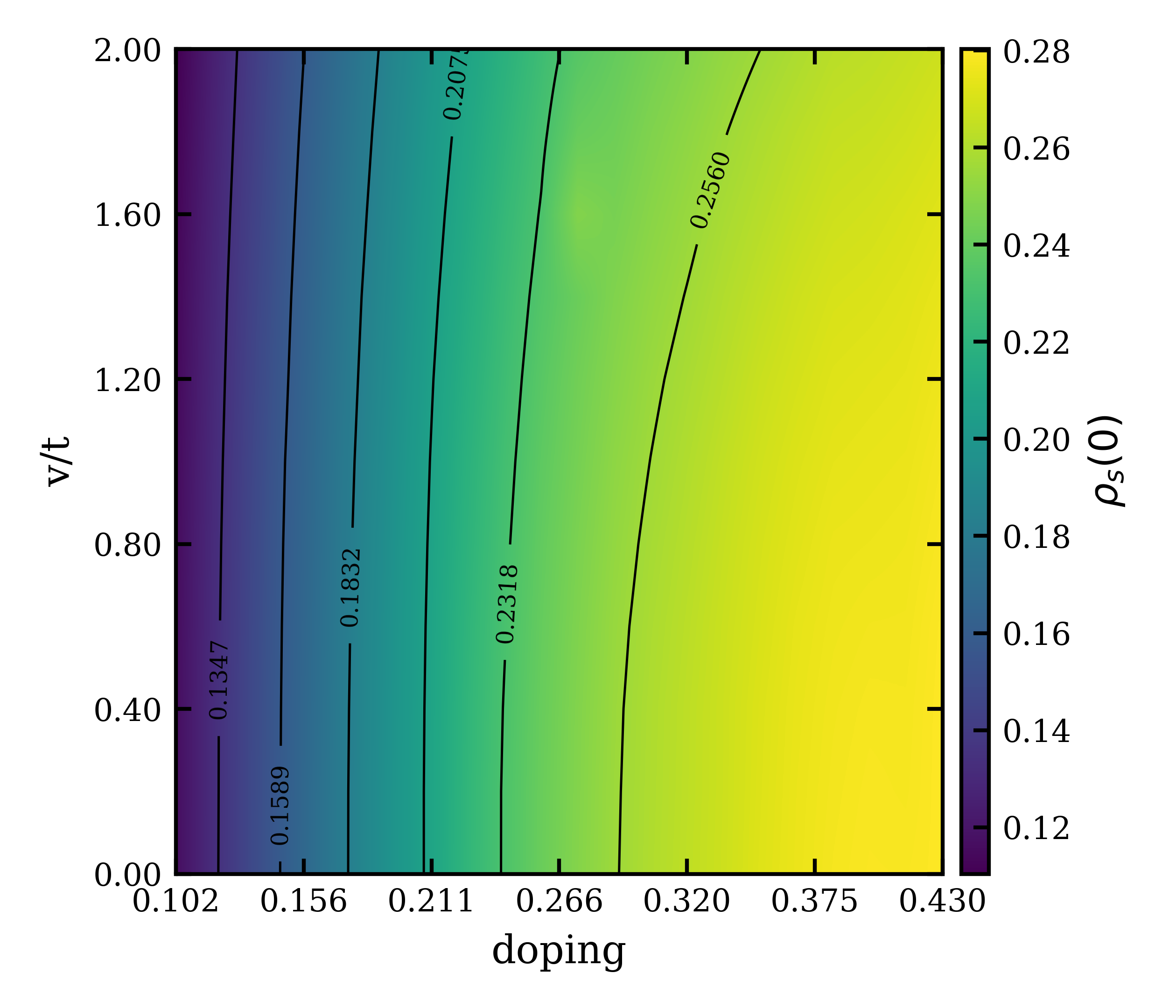}
\caption{The doping-disorder strength evolution of $\rho_{s}$ computed with the two-component wave function Eq.36(upper panel) and its generalization Eq.41(lower panel) for the disordered 2D $t-J$ model. The calculation is done on a $16\times16$ lattice with periodic boundary condition in the $x$-direction and anti-periodic boundary condition in the $y$-direction.}
\end{figure}  

We find that the evolution of $\rho_{s}(0)$ follows closely that of the total optical weight $K$, whose full doping-disorder strength evolution is illustrated in Fig.4. We find that a substantial part of the total optical weight remain un-condensed even at zero temperature for the disordered $t-J$ model. The ratio of the condensed part over the total optical weight is illustrated in Fig.5. It is found that such a ratio increase monotonically with the doping concentration but is still substantially smaller than 1 at large doping. Such strong electron incoherence should mainly be attributed to the strong correlation effect in the $t-J$ model rather than the disorder effect, since the ratio $r=\rho_{s}/K$ is seen to exhibit only a very gentle dependence on the disorder strength $V$. Thus, unlike the situation in a disordered BCS superconductor, the strong electron incoherence in the disordered $t-J$ model is largely an intrinsic effect caused by the strong correlation between the electron, rather than a disorder induced extrinsic effect.

\begin{figure}
\includegraphics[width=8cm]{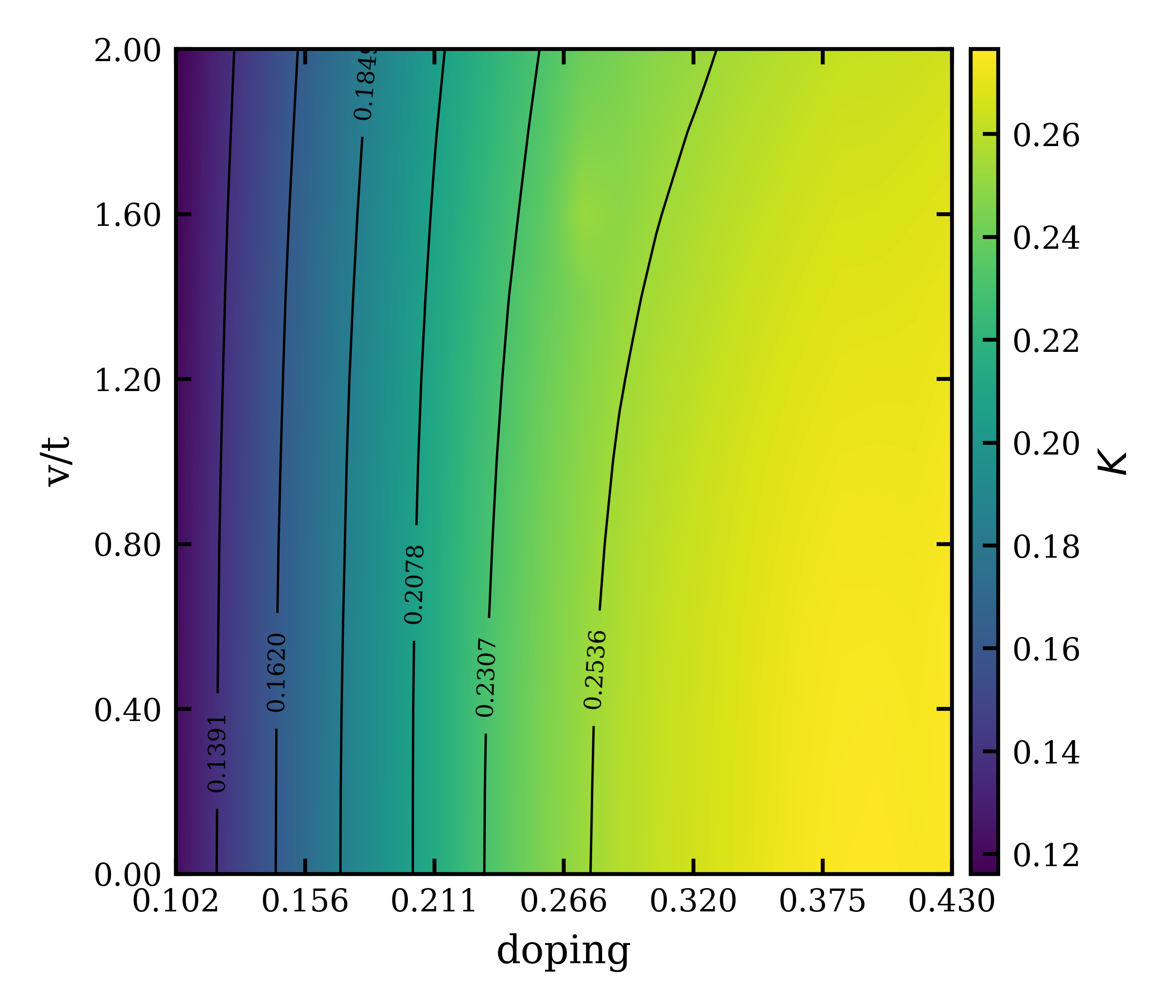}
\includegraphics[width=8cm]{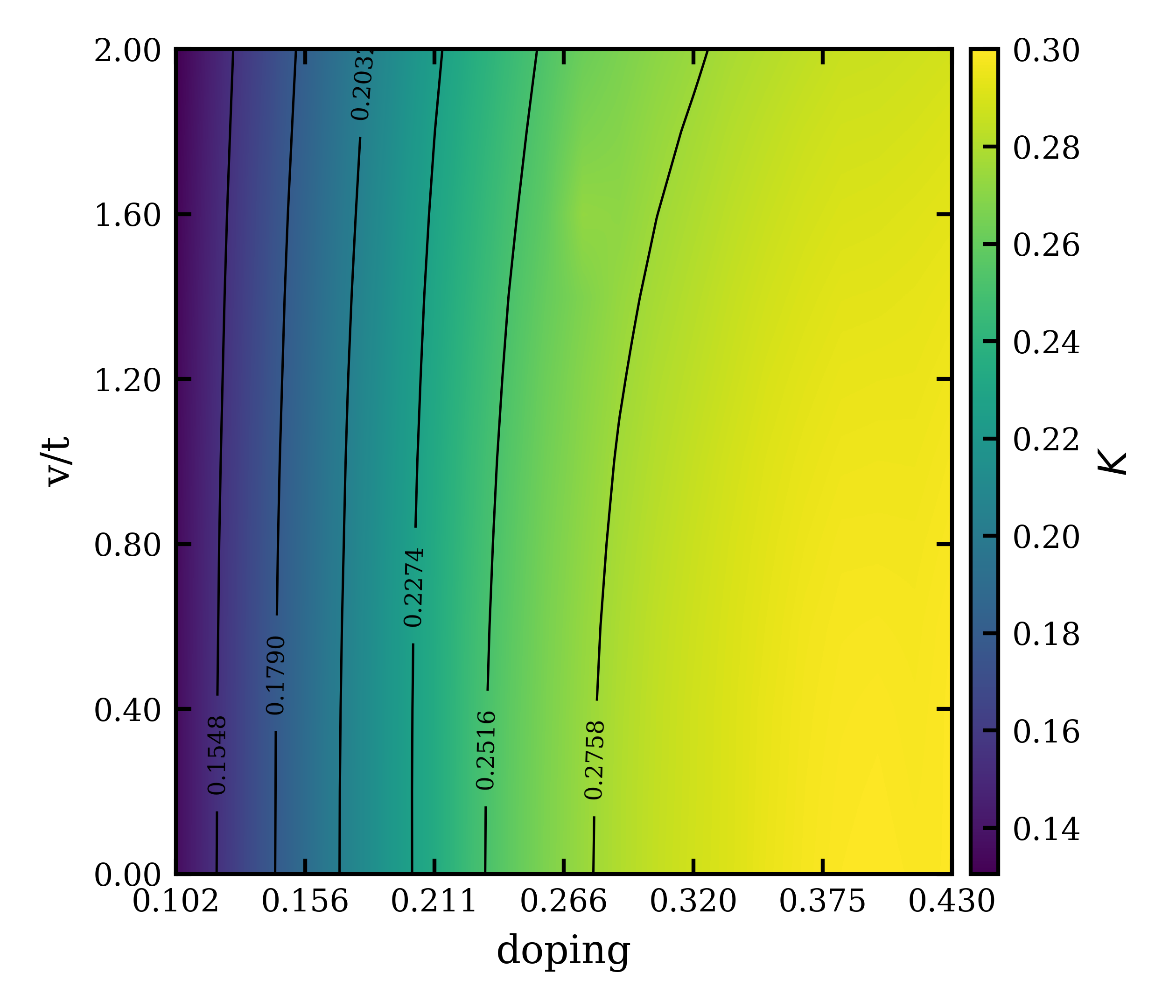}
\caption{The doping-disorder strength evolution of the total optical weight $K$ computed with the two-component wave function Eq.36(upper panel) and its generalization Eq.41(lower panel) for the disordered 2D $t-J$ model. The calculation is done on a $16\times16$ lattice with periodic boundary condition in the $x$-direction and anti-periodic boundary condition in the $y$-direction.}
\end{figure}  

\begin{figure}
\includegraphics[width=8cm]{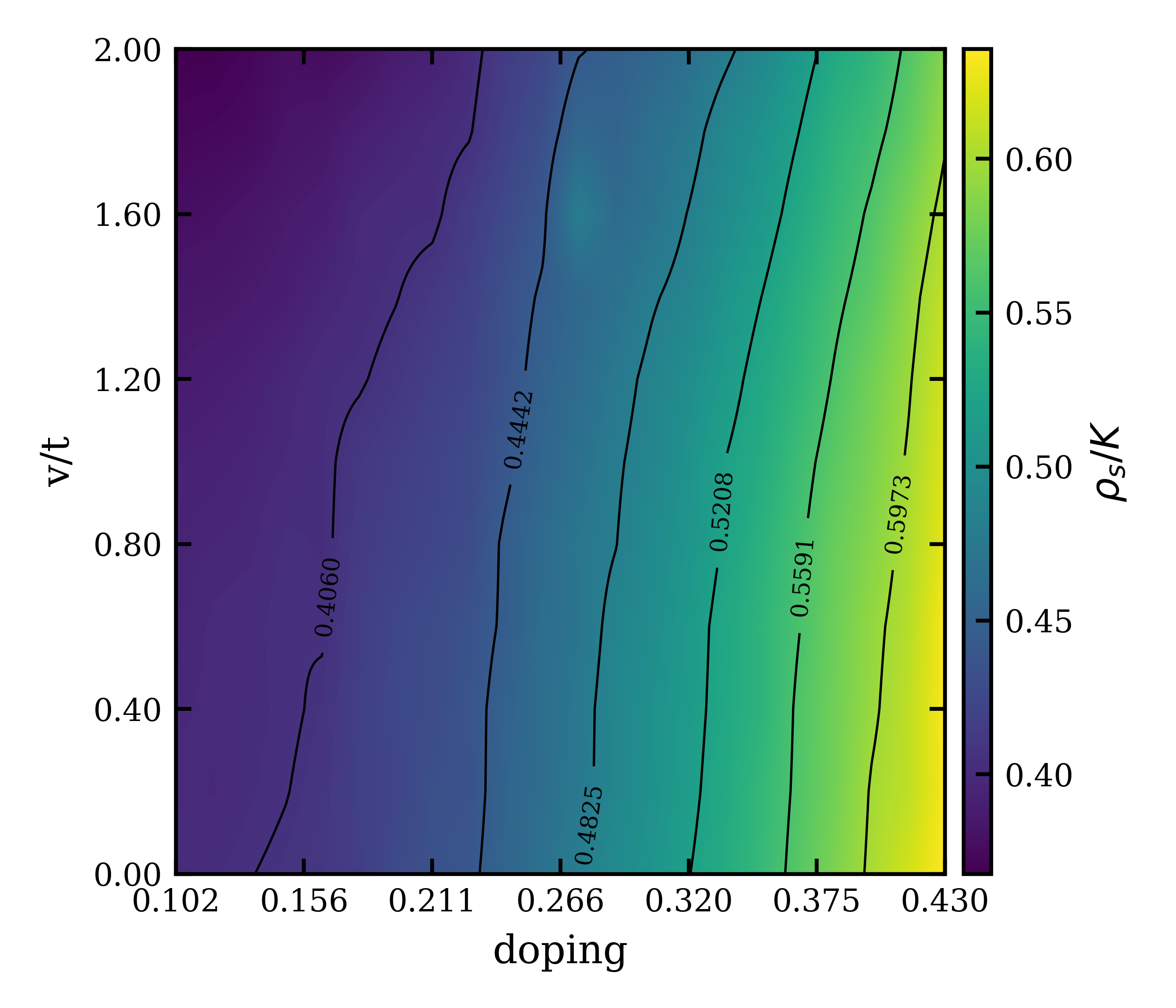}
\includegraphics[width=8cm]{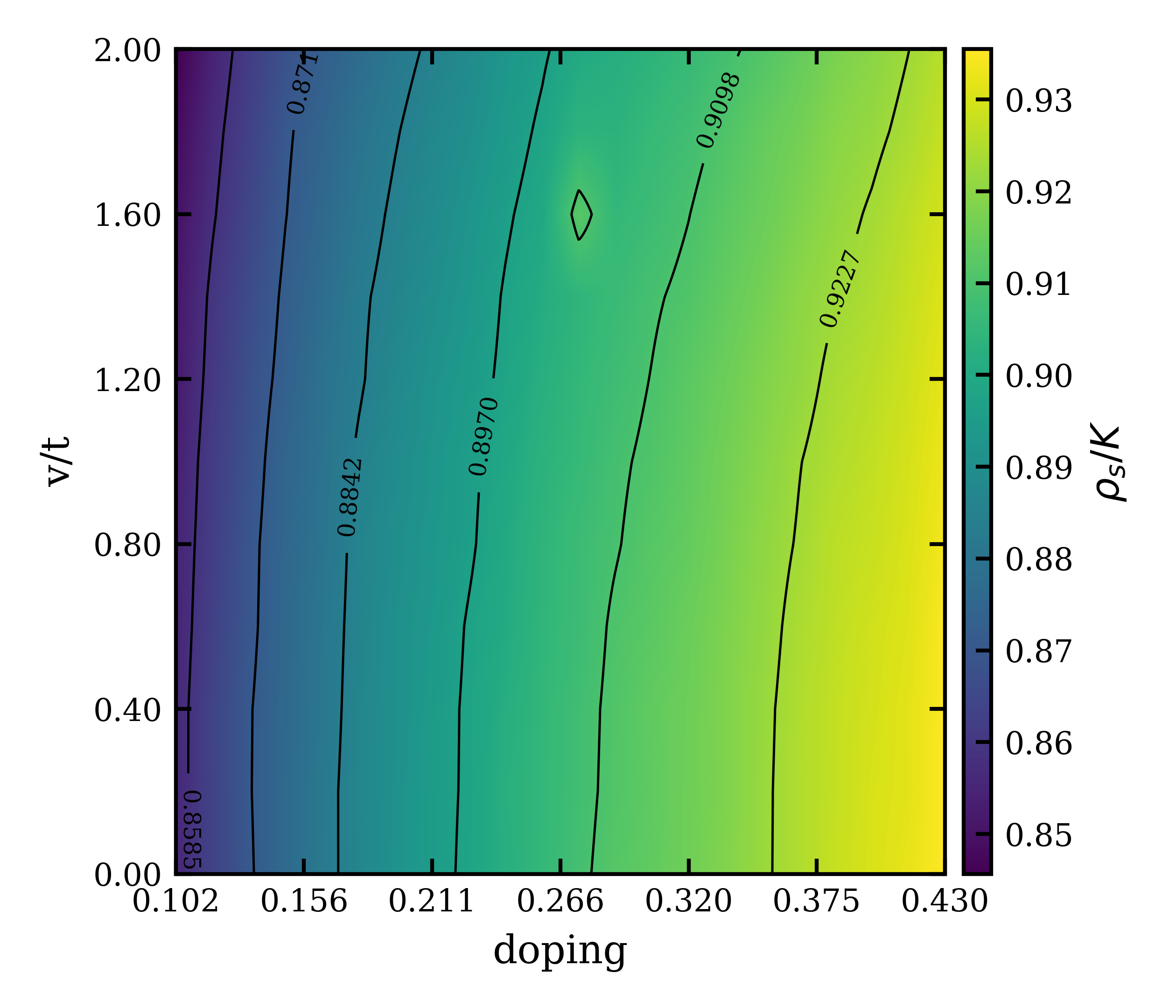}
\caption{The doping-disorder strength evolution of the ratio $r=\rho_{s}(0)/K$ computed with the two-component wave function Eq.36(upper panel) and its generalization Eq.41(lower panel) for the disordered 2D $t-J$ model. The calculation is done on a $16\times16$ lattice with periodic boundary condition in the $x$-direction and anti-periodic boundary condition in the $y$-direction.}
\end{figure}  

It is interesting to note that the intrinsic electron incoherence related to the strong correlation effect actually acts to protect the d-wave superconductivity in the $t-J$ model from the suppression of the disorder effect. In other words, the intrinsic electron incoherence preempts the chance for the disorder potential to suppress the d-wave superconductivity in the system. This may answer the general question why high temperature superconductivity is found in such a strongly correlated background rather than in a less correlated metal. In our previous study\cite{Yang}, we propose that such a protection can be more accurately ascribed to the spin-charge separation mechanism in a doped Mott insulating, or quantum spin liquid state. Indeed, spin-charge separation provides a natural way to realize strong electron incoherence in the $t-J$ model.    

\subsection{A comparison with the prediction of the BCS mean field theory}
To illustrate the role of the strong correlation effect further, we now present the result of $\rho_{s}(0)$ computed from the BCS mean field approximation for the disordered $t-J$ model. To maintain a comparable size in the paring gap as that predicted by the RVB theory, here we increase deliberately the strength of the Heisenberg exchange coupling to a value of $J=t$. The evolution of the $\rho_{s}(0)$ with the disorder strength at $x=0.236$ is plotted in Fig.6. This is to be compared with the result computed from the RVB theory in the last subsection. Indeed, the superfluidity in the BCS mean field theory is much more fragile than that predicted by the RVB theory.  

\begin{figure}
\includegraphics[width=8cm]{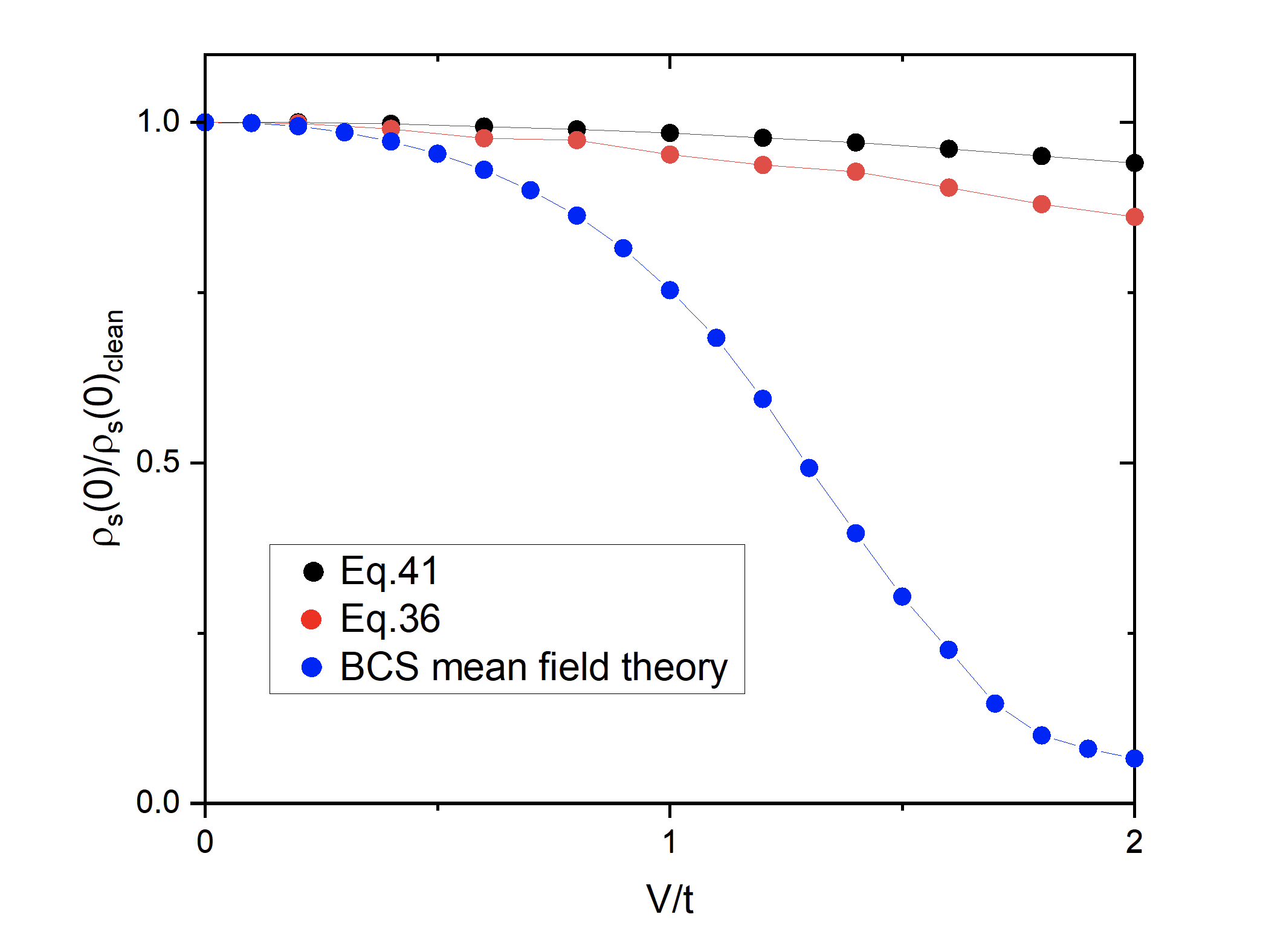}
\caption{Comparison between the zero temperature superfluid density computed from the RVB theory and the BCS mean field theory. Plotted here is the zero temperature superfluid density $\rho_{s}(0)$ normalized by its clean limit value, which is denoted here as $\rho_{s}(0)_{clean}$. The BCS mean field calculation is performed on a $6400\times6400$ lattice with the supercell trick. Here each supercell is of size $32\time32$ and there are $200\times200$ such supercells tiled periodically in both the $x$-direction and the $y$-direction. The RVB result for the normalized zero temperature superfluid density computed with the two-component wave function Eq.36 and Eq.41 are very similar to each other. The calculation for the RVB theory is done on a $16\times16$ lattice with periodic boundary condition in the $x$-direction and anti-periodic boundary condition in the $y$-direction.}
\end{figure}  

Finally, we note that the ratio between the zero temperature superfluid density and the total optical weight, namely $r=\rho_{s}(0)/K$ introduced in the last subsection is rapidly suppressed 
with the increase of the disorder strength in the BCS mean field theory. According to BCS theory, $r$ should approach 1 in the clean limit, implying that all itinerant spectral weight should go to the superconducting condensate at zero temperature in the clean limit. The strong reduction of $r$ from 1 means that a substantial part of the optical weight would remain incoherent at zero temperature in the presence of strong disorder. This is indeed what is observed in heavily overdoped cuprate. However, such a behavior is in strong contrast with the result we obtained from the RVB theory, according to which $r=\rho_{s}(0)/K$ only exhibits a gentle dependence on the disorder strength.

\section{Conclusions and discussions}

In this work, we develop a variational theory for the zero temperature superfluid density $\rho_{s}(0)$ of the disordered $t-J$ model. This is achieved in two steps. First, we perform an unrestricted variational optimization of a fermionic RVB state for the ground state of the disordered $t-J$ model. We then construct the variational state that can describe the paramagnetic current response on such an RVB state in the presence of the external EM field. The zero temperature superfluid density $\rho_{s}(0)$ is extracted from the curvature of the variational ground state energy of the system as a function of the external EM field.

We find that the zero temperature superfluid density computed in this way exhibits remarkable robustness against the disorder potential. More specifically, we find that $\rho_{s}(0)$ is a monotonically increasing function of doping concentration $x$ and it exhibits only a rather weak dependence on the disorder strength. The doping and the disorder strength evolution of $\rho_{s}(0)$ is found to follow closely those of the total optical weight $K$. These results are qualitatively consistent with the observations in the underdoped regime, in which that $\rho_{s}(0)$ is found to track closely the doping dependence of the Drude weight. We find that there is substantial electron incoherence even in the absence of the disorder potential. The computed ratio $r=\rho_{s}(0)/K$ is found to have only a rather weak dependence on the disorder strength.  Clearly, such strong electron incoherence should be attributed to strong correlation effect between the electron. Interestingly, such strong intrinsic electron incoherence acts to protect the d-wave superconductivity from the suppression by the disorder effect. This is consistent with the observation in our previous study, in which it is found that the d-wave spinon pairing is protected by the spin-charge separation mechanism in a doped spin liquid background. 

We have also performed a BCS mean field calculation of $\rho_{s}(0)$ by ignoring the no double occupancy constraint in the $t-J$ model. This is believed to be relevant in the overdoped cuprates. Consistent with previous studies adopting such a weak coupling approach, we find that $\rho_{s}(0)$ is quickly suppressed with the increase of the disorder strength. The ratio $\rho_{s}(0)/K$, which approaches $1$ in the clean limit, is found to be rapidly suppressed with the increase of the disorder strength. This results in large amount of spectral weight that fail to condense even at zero temperature. This is consistent with the observation in the heavily overdoped cuprates\cite{Bozovic,Mahmood}.

The strong contrast between the underdoped and the overdoped cuprates in the disorder effect lends strong support to our previous proposal that there exists a Mott transition at the pseudogap end point across which a strongly correlated doped Mott insulating metal is transmuted into a weakly correlated fermi liquid metal\cite{Yang,Sordi}. No symmetry breaking is involved in such a non-Landau transition. The superfluidity in the doped Mott insulating phase gains its robustness against the disorder effect from the intrinsic electron incoherence in such a strongly correlated system. On the other hand, the d-wave superconductivity in the overdoped cuprate becomes fragile against disorder effect without such a protection mechanism. We believe that the dramatic strange metal behavior observed observed around the pseudogap end point should be attributed to such a Mott transition.

Phenomenologically, there are many evidences implying a rapid transition of the electronic state around the pseudogap end point. For example, ARPES measurements find that the peak-dip-hump structure in the anti-nodal spectrum is suddenly suppressed across the pseudogap end point\cite{Chen}. This is accompanied by the RIXS observation that the itinerant nature of the spin fluctuation is suddenly enhanced around the same critical doping\cite{Minola}. In addition, $x^{*}$ is also found to be position where field induced spin glass behavior starts to emerge\cite{Julien}. The study of the robustness of the superfluidity against the disorder effect as attempted here provides an additional opportunity to strengthen such a scenario. We hope that this work would stimulate experimental interest in a deeper understanding of the disorder effect in such a strongly correlated electron system, in particular, in system with controllable level of disorder. We also note that the variational approach adopted in this work is at the best of semi-quantitative nature. Study of the disorder effect in the $t-J$ model with more advanced numerical approaches is not only possible and urgent, but must be fruitful.

\begin{acknowledgments}
Rong Cheng and Tao Li acknowledge the support from the National Natural Science Foundation of China(Grant No.12274457), Jianhua Yang acknowledge the support from the National Natural Science Foundation of China(Grant No.12504178).
\end{acknowledgments}

\appendix
\section{The relation between the RVB mean field ansatz $H^{f}_{MF}$ and the RVB amplitude $a(i,j)$}
In section II, we mentioned that an RVB amplitude $a(i,j)$ can be uniquely generated from a RVB mean field ansatz of the form Eq.20. Here we derive the explicit form for $a(i,j)$. For this purpose, we rewrite the mean field ansatz Eq.20 in the following matrix form
\begin{equation}
H^{f}_{\mathrm{MF}}=\bm{\psi}^{\dagger}\mathbf{M}\bm{\psi}
\end{equation}
here $\bm{\psi}$ is given by
\begin{equation}
\bm{\psi}^{\dagger}=(f^{\dagger}_{1,\uparrow},...,f^{\dagger}_{N,\uparrow},f_{1,\downarrow},....,f_{N,\downarrow})
\end{equation}
$\mathbf{M}$ is a $2N\times 2N$ Hermitian matrix of the form
\begin{equation}
\mathbf{M}=-\left(\begin{array}{cc} \chi_{i,j} & \Delta_{i,j} \\ \Delta^{*}_{i,j} & -\chi_{j,i} \end{array}\right)+\left(\begin{array}{cc}\mu_{i}\delta_{i,j} & 0 \\ 0 & -\mu_{i}\delta_{i,j} \end{array}\right)
\end{equation}
We will not specify the detailed relation between the parameters $\chi_{i,j}$, $\Delta_{i,j}$ and $\mu_{i}$ appearing here and the parameters $t^{f}_{i,j}$, $t'^{f}_{i,j}$, $\Delta^{f}_{i,j}$ and $\mu^{f}_{i}$ appearing in Eq.20. However, we will assume all these parameters to be real numbers. This is sufficient for our purpose, since the RVB state we optimize is time reversal invariant when the external EM field is zero.
 
$H^{f}_{\mathrm{MF}}$ can be brought into the following diagonalized form under unitary transformation 
 \begin{equation}
H^{f}_{\mathrm{MF}}=\sum_{i=1}^{N}\epsilon_{i}(\gamma^{\dagger}_{i,\uparrow}\gamma_{i,\uparrow}-\gamma_{i,\downarrow}\gamma^{\dagger}_{i,\downarrow})
\end{equation}
here
\begin{equation}
\bm{\gamma}^{\dagger}=(\gamma^{\dagger}_{1,\uparrow},...,\gamma^{\dagger}_{N,\uparrow},\gamma_{1,\downarrow},....,\gamma_{N,\downarrow})
\end{equation}
is related to $\bm{\psi}$ by the following unitary transformation 
\begin{equation}
\bm{\psi}=\left(\begin{array}{cc}\mathbf{u} & \mathbf{v} \\-\mathbf{v} & \mathbf{u}\end{array}\right)\bm{\gamma}
\end{equation} 
in which $\mathbf{u}$ and $\mathbf{v}$ are $N\times N$ real matrix, $\epsilon_{i}$ is the $i$-th eigenvalue of matrix $\mathbf{M}$. We note that as a result of the spin rotational symmetry, these eigenvalues always appear in $\pm$ pairs. Here we assume that the first $N$ eigenvalues are positive. 

If we denote the mean field ground state of $H^{f}_{\mathrm{MF}}$ as $|\mathrm{MF}\rangle$, then it must satisfy
\begin{equation}
\gamma_{i,\uparrow}|\mathrm{MF}\rangle=\gamma_{i,\downarrow}|\mathrm{MF}\rangle=0
\end{equation}
It can then be shown that
\begin{equation}
|\mathrm{MF}\rangle\propto e^{\sum_{i,j}a(i,j)f^{\dagger}_{i,\uparrow}f^{\dagger}_{j,\downarrow}}|0\rangle
\end{equation}
in which
\begin{equation}
a(i,j)=(\mathbf{u}^{-1}\mathbf{v})_{i,j}
\end{equation}
is the wave function of the spinon Cooper pair in the mean field theory, or, the required RVB amplitude.


\begin{thebibliography}{}

\bibitem{Proust}C. Proust and L. Taillefer, "The remarkable underlying ground states of cuprate superconductors", Ann. Rev. Cond. Matt. Phys. \textbf{10}, 409 (2019). 
\bibitem{Hussey}N. E. Hussey, J. Buhot, S. Licciardello, "A tale of metals: contrasting criticalities in the pnictides and hole-doped cuprates", Rev. Prog. Phys. \textbf{81}, 052501 (2018).
\bibitem{Chen}S. D. Chen,  M. Hashimoto, Y. He, D. J. Song, K. J. Xu, J. F. He, T. P. Devereaux, H. Eisaki, D. H. Lu, J. Zaanen, and Z. X. Shen, "Incoherent strange metal sharply bounded by a critical doping in Bi2212", Science, \textbf{366},1099(2019).

\bibitem{Taillefer}L. Taillefer, "Scattering and pairing in cuprate superconductors", Ann. Rev. Cond. Matt. Phys. \textbf{1}, 50 (2010).

\bibitem{CV}B. Michon, C. Girod, S. Badoux, J. Ka\v{c}mar\v{c}\'{i}k, Q. Ma, M. Dragomir, H. A. Dabkowska, B. D. Gaulin, J.-S. Zhou, S. Pyon, T. Takayama, H. Takagi, S. Verret, N. Doiron-Leyraud, C. Marcenat, L. Taillefer, T. Klein, "Thermodynamic signatures of quantum criticality in cuprates", Nature \textbf{567}, 218 (2019).

\bibitem{Hall3} S. Badoux, W. Tabis, F. Laliberte, G. Grissonnanche, B. Vignolle, D. Vignolles, J. Beard, D. A. Bonn, W. N. Hardy, R. Liang, N. Doiron-Leyraud, L. Taillefer, and C. Proust, Change of carrier density at the pseudogap critical point of a cuprate superconductor, Nature (London) \textbf{531}, 210 (2016).
\bibitem{Tallon}J.L. Tallon and J.W. Loram, "The doping dependence of T$^{*}$ – what is the real high-T$_{c}$ phase diagram?", Physica C \textbf{349}, 53 (2001).

\bibitem{Uemura}Y. J. Uemura, G. M. Luke, B. J. Sternlieb, J. H. Brewer, J. F. Carolan, W. N. Hardy, R. Kadono, J. R. Kempton, R. F. Kiefl, S. R. Kreitzman, P. Mulhern,
T. M. Riseman, D. L. Williams, B. X. Yang, S. Uchida, H. Takagi, J. Gopalakrishnan, A. W. Sleight, M. A. Subramanian, C. L. Chien, M. Z. Cieplak, G. Xiao, V. Y. Lee, B. W. Statt, C. E. Stronach, W. J. Kossler, X. H. Yu, "Universal correlations between T$_{c}$ and $\frac{n_{s}}{m^{*}}$ (carrier density over effective mass) in high-T$_{c}$ cuprate superconductors", Phys. Rev. Lett. \textbf{62}, 2317(1989).
\bibitem{Bozovic}I. Bo\v{z}ovi\'{c}, X. He, J. Wu and A. T. Bollinger, "Dependence of the critical temperature in overdoped copper oxides on superfluid density", Nature,\textbf{536}, 309(2016).
\bibitem{Michon}B. Michon, A.B. Kuzmenko, M.K. Tran, B. McElfresh, S. Komiya, S. Ono, S. Uchida and D. van der Marel, "Spectral weight of hole-doped cuprates across the pseudogap critical point", Phys. Rev. Research \textbf{3}, 043125 (2021).

\bibitem{Hone1}N. R. Lee-Hone, J. S. Dodge, and D. M. Broun, "Disorder and superfluid density in overdoped cuprate superconductors", Phys. Rev. B \textbf{96}, 024501 (2017).
\bibitem{Hone}N. R. Lee-Hone, H. U. \"{O}zdemir, V. Mishra, D. M. Broun, and P. J. Hirschfeld, "Low energy phenomenology of the overdoped cuprates: Viability of the Landau-BCS paradigm", Phys. Rev. Research \textbf{2}, 013228 (2020).
\bibitem{Lizx} Z. X. Li, S. A. Kivelson, and D. H. Lee, "Superconductor-to-metal transition in overdoped cuprates", npj Quantum Materials \textbf{6}, 36 (2021).
\bibitem{Kivelson}B. J. Ramshaw and S. A. Kivelson, "Superconductivity in overdoped cuprates can be understood from a BCS perspective!", arXiv:2510.25767.
\bibitem{Broun}D. M. Broun, H. U. \"{O}zdemir, V. Mishra, N. R. Lee-Hone, X. Kong, T. Berlijn, and P.J.Hirschfeld, "Optical conductivity of overdoped cuprates from ab-initio out-of-plane impurity potentials", Phys. Rev. B, \textbf{109}, 174519(2024).
\bibitem{Berg}J. Kim, E. Berg, and E. Altman, "Theory of a strange metal in a quantum superconductor to metal transition", arXiv:2401.17353.

\bibitem{Markowitz}D. Markowitz and L. P. Kadanoff, "Effect of impurities upon critical temperature of anisotropic superconductors", Phys. Rev. \textbf{131}, 563(1963).
\bibitem{Alloul} H. Alloul, J. Bobroff, M. Gabay, and P. J. Hirschfeld, "Defects in correlated metals and superconductors", Rev. Mod. Phys. \textbf{81}, 45 (2009).
\bibitem{Atkinson} W. A. Atkinson, P. J. Hirschfeld, and A. H. MacDonald, "Gap inhomogeneities and the density of states in disordered $d$-wave superconductors", Phys. Rev. Lett. \textbf{85} 3922(2000).
\bibitem{Xiang}T. Xiang and J. M. Wheatley, "Nonmagnetic impurities in two-dimensional superconductors", Phys. Rev. B \textbf{51}, 11721 (1995).
\bibitem{Atkinson1}W. A. Atkinson and P. J. Hirschfeld, "Optical and thermal transport properties of an inhomogeneous $d$-wave superconductor", Phys. Rev. Lett. \textbf{88} 187003(2002).
\bibitem{Anderson}P. W. Anderson, "Theory of dirty superconductors", J. Phys. Chem. Solids \textbf{11}, 26 (1959).

\bibitem{Mahmood}F. Mahmood, X. He, I. Bo\v{z}ovi\'{c} and N. P. Armitage, "Locating the missing superconducting electrons in the overdoped cuprates La$_{2-x}$Sr$_{x}$CuO$_{4}$" Phys. Rev. Lett., \textbf{122}, 027003(2019).

\bibitem{Tromp}W. O. Tromp, T. Benschop, J. F. Ge, I. Battisti, K. M. Bastiaans, D. Chatzopoulos, A. H. M. Vervloet, S. Smit, E. van Heumen, M. S. Golden, Y. Huang, T. Kondo, T. Takeuchi, Y. Yin, J. E. Hoffman, M. Antonio Sulangi, J. Zaanen and M. P. Allan, "Puddle formation and persistent gaps across the non-mean-field breakdown of superconductivity in overdoped (Pb,Bi)$_{2}$Sr$_{2}$CuO$_{6+\delta}$", Nature Materials \textbf{22}, 703(2023).


\bibitem{Grag}A. Garg, M. Randeria, and N. Trivedi, "Strong correlations make high-temperature superconductors robust against disorder", Nat. Phys. \textbf{4}, 762 (2008).
\bibitem{Chakra}D. Chakraborty and A. Ghosal, "Fate of disorder-induced inhomogeneities in strongly correlated d-wave superconductors", New Journal of Physics \textbf{16}, 103018 (2014).
\bibitem{Ghosal}D. Chakraborty, N. Kaushal, and A. Ghosal, "Pairing theory for strongly correlated $d$-wave superconductors", Phys. Rev. B \textbf{96}, 134518 (2017).

\bibitem{Yang}Jianhua Yang and Tao Li, "Origin of the pseudogap end point in the high-T$_{c}$ cuprate superconductors", Phys. Rev. B \textbf{110}, 024521 (2024).

 \bibitem{Scalapino}D. J. Scalapino, S. R. White and S. C. Zhang, "Insulator, metal, or superconductor: The criteria", Phys. Rev. B \textbf{47}, 7995(1993).
 
 \bibitem{RVB}P. W. Anderson, "The Resonating Valence Bond State in La$_{2}$CuO$_{4}$ and Superconductivity", Science \textbf{235}, 1196(1987).
\bibitem{PALee}P. A. Lee, N. Nagaosa and X. G. Wen, "Doping a Mott insulator: Physics of high-temperature superconductivity", Rev. Mod. Phys. \textbf{78}, 17 (2006).

\bibitem{Li} Jian-Hua Yang and Tao Li, "Instability of the $U(1)$ spin liquid with a large spinon Fermi surface in the Heisenberg-ring exchange model on the triangular lattice", Phys. Rev. B, \textbf{108} 235105(2023).

\bibitem{Li2} Jian-Hua Yang and Tao Li, "Strong relevance of Zinc impurity in the spin-$\frac{1}{2}$ Kagome quantum antiferromagnets: a variational study", Phys. Rev. B, \textbf{109} 115103(2024).

\bibitem{Rong1}Rong Cheng and Tao Li, "Closely competing valence bond crystal orders in the ground state of the spin-$\frac{1}{2}$ antiferromagnetic Heisenberg model on the pyrochlore lattice: a large scale unrestricted variational study", Phys. Rev. B, \textbf{113} 075158(2026).

\bibitem{Rong2}Rong Cheng and Tao Li, "Variational study of the magnetization plateaus of the spin-$\frac{1}{2}$ kagome Heisenberg antiferromagnet and its implication on YCOB", Phys. Rev. B, \textbf{113} 085136(2026).

\bibitem{Li1}Tao Li and Fan Yang, "Variational study of the neutron resonance mode in the cuprate superconductors", Phys. Rev. B \textbf{81}, 214509 (2010).

\bibitem{Sordi}G. Sordi, P. S\'{e}mon, K. Haule, and A.-M. S. Tremblay, "Strong coupling superconductivity, pseudogap, and Mott transition", Phys. Rev. Lett. \textbf{108}, 216401(2012).

\bibitem{Minola}M. Minola, Y. Lu, Y. Y. Peng, G. Dellea, H. Gretarsson, M. W. Haverkort, Y. Ding, X. Sun, X. J. Zhou, D. C. Peets, L. Chauviere, P. Dosanjh, D. A. Bonn, R. Liang, A. Damascelli, M. Dantz, X. Lu, T. Schmitt, L. Braicovich, G. Ghiringhelli, B. Keimer and M. Le Tacon,  "Crossover from collective to incoherent spin excitations in superconducting cuprates probed by detuned resonant inelastic X-ray scattering", Phys. Rev. Lett., \textbf{119}, 097001(2017).

\bibitem{Julien}M. Frachet, I. Vinograd, R. Zhou, S. Benhabib, S. Wu, H. Mayaffre, S. Kr\"{a}mer, S. K. Ramakrishna, A. P. Reyes, J. Debray, T. Kurosawa, N. Momono, M. Oda, S. Komiya, S. Ono, M. Horio, J. Chang, C. Proust, D. LeBoeuf, M. Julien, "Hidden magnetism at the pseudogap critical point of a cuprate superconductor", Nat. Phys., \textbf{16}, 1064(2020). 
  

\end{thebibliography}
\end{document}